\documentclass[a4paper,preprintnumbers,showpacs,onecolumn,superscriptaddress,nofootinbib,amsmath,amssymb,notitlepage]{revtex4-2}
\usepackage{mathtools,leftindex,tensor,mhchem}
\usepackage{booktabs}
\usepackage{siunitx}
\usepackage{enumitem}
\usepackage{times}
\usepackage{dcolumn}
\usepackage{mathrsfs}
\usepackage{comment}
\usepackage{float}
\usepackage{hyperref}
\usepackage{amsmath,accents}
\usepackage{mathtools}
\usepackage{bbm}
\usepackage{bm}
\usepackage{amsfonts}
\usepackage{amssymb}
\usepackage{mathrsfs}
\usepackage[scr=rsfs,cal=boondox]{mathalfa}
\usepackage{wasysym}
\usepackage{graphicx}
\usepackage{filecontents}
\usepackage[dvipsnames]{xcolor}
\usepackage{bbold}
\usepackage[scr=rsfs,cal=boondox]{mathalfa}
\usepackage{placeins}
\usepackage{subcaption}

\usepackage{stackengine}
\stackMath
\newcommand\tsup[2][2]{%
 \def\useanchorwidth{T}%
  \ifnum#1>1%
    \stackon[-1.3ex]{\tsup[\numexpr#1-1\relax]{#2}}{\mathchar"307E}%
  \else%
    \stackon[-1ex]{#2}{\mathchar"307E}%
  \fi%
}
\hypersetup{
    bookmarks=true,         % show bookmarks bar?
    pdftoolbar=true,        % show Acrobats toolbar?
    pdfmenubar=true,        % show Acrobats menu?
    pdffitwindow=true,     %window fit to page when opened
    pdfstartview={FitH},     %fits the width of the page to the window
    pdftitle={My title},     %title
    pdfauthor={author},      %author
    pdfsubject={Subject},    %subject of the document
    pdfcreator={Creator},    %creator of the document
    pdfproducer={Producer},  %producer of the document
    pdfkeywords={keyword1} %{key2} {key3}, % list of keywords
    pdfnewwindow=true,       %links in new window
    colorlinks=true ,        %false: boxed links; true: colored links
    linkcolor=Blue  ,        %color of internal links (change box color with linkbordercolor)
    citecolor=Blue,      %color of links to bibliography
    filecolor=Blue,       %color of file links
    urlcolor=Blue            %color of external links
}

\newcommand{\ed}{\mathrm{d}}

\newcommand{\mL}{\mathcal{L}}
\newcommand{\mc}{\mathcal{c}}

\newcommand{\oalpha}[1]{\accentset{\circ}{\alpha}}
\newcommand{\obf}[1]{\accentset{\circ}{\mathbf{f}}}
\newcommand{\boR}[1]{\accentset{\circ}{\mathbf{R}}}
\newcommand{\obF}[1]{\accentset{\circ}{\mathbf{F}}}
\newcommand{\obPi}[1]{\accentset{\circ}{\mathbf{\Pi}}}

\usepackage{scalerel}
\usepackage{tikz}
\usetikzlibrary{svg.path}

\definecolor{orcidlogocol}{HTML}{A6CE39}
\tikzset{
  orcidlogo/.pic={
    \fill[orcidlogocol] svg{M256,128c0,70.7-57.3,128-128,128C57.3,256,0,198.7,0,128C0,57.3,57.3,0,128,0C198.7,0,256,57.3,256,128z};
    \fill[white] svg{M86.3,186.2H70.9V79.1h15.4v48.4V186.2z}
                 svg{M108.9,79.1h41.6c39.6,0,57,28.3,57,53.6c0,27.5-21.5,53.6-56.8,53.6h-41.8V79.1z M124.3,172.4h24.5c34.9,0,42.9-26.5,42.9-39.7c0-21.5-13.7-39.7-43.7-39.7h-23.7V172.4z}
                 svg{M88.7,56.8c0,5.5-4.5,10.1-10.1,10.1c-5.6,0-10.1-4.6-10.1-10.1c0-5.6,4.5-10.1,10.1-10.1C84.2,46.7,88.7,51.3,88.7,56.8z};
  }
}

\newcommand\orcidicon[1]{\href{https://orcid.org/#1}{\mbox{\scalerel*{
\begin{tikzpicture}[yscale=-1,transform shape]
\pic{orcidlogo};
\end{tikzpicture}
}{|}}}}
\begin{document}

%%%%%%
\title{Thermodynamic and observational constraints on black holes with primary hair in Beyond Horndeski Gravity: Stability and shadows}

\author{Cristian Erices\orcidicon{0000-0001-5852-514X}}
    \email{cristian.erices@ucentral.cl}
\affiliation{
Facultad de Ingeniería y Arquitectura, Universidad Central de Chile,
Av. Santa Isabel 1186, 8330563, Santiago, Chile}
%\affiliation{Grupo de Ciencias del Espacio y Físicas, Universidad Central de Chile, Toesca 1783, Santiago 8320000, Chile}
\affiliation{Departamento de Matemática, Física y Estadística, Universidad Católica del Maule, Av. San Miguel 3605, Talca 3480094, Chile}

\author{Mohsen Fathi\orcidicon{0000-0002-1602-0722}}
\email{mohsen.fathi@ucentral.cl}
\affiliation{
Facultad de Ingeniería y Arquitectura, Universidad Central de Chile,
Av. Santa Isabel 1186, 8330563, Santiago, Chile}
%\affiliation{Grupo de Ciencias del Espacio y Físicas, Universidad Central de Chile, Toesca 1783, Santiago 8320000, Chile}

%%%%%%%%%%%%%%%%%%%abstract
\begin{abstract}

In this paper, we find that unlike in General Relativity, the shift-symmetric subclass of Beyond Horndeski theories permits black holes with primary hair that are thermodynamically stable and align with current Event Horizon Telescope observations of the M87* and Sgr A* black holes. This work begins by investigating thermodynamic properties, analyzing how primary hair influences thermodynamic quantities and local stability, which imposes strict constraints on the allowed range of primary hair values. The null geodesics near this black hole are then examined, demonstrating how scalar hair affects the shadow diameter. Specifically, when the parameter of the Beyond Horndeski function $F_4$ is negative, increasing scalar hair enlarges the shadow; in contrast, when this parameter is positive, greater scalar hair reduces the shadow size. Further constraints on the scalar hair are derived using observational data, highlighting its sensitivity to other black hole parameters. To explore additional observational features, face-on two-dimensional images of spherically infalling accretion disks are simulated, revealing how primary scalar hair shapes the black hole's shadow. Finally, all relevant constraints are combined to identify black holes that are both stable and consistent with observational data.

%We found that, unlike General Relativity, the shift-symmetric subclass of Beyond Horndeski theories provides black holes with primary hair, which are thermodynamically stable and compatible with current Event Horizon Telescope observations for M87* and Sgr A* black holes. We first investigate its thermodynamic properties by studying the influence of the primary hair on the thermodynamic quantities and the local stability, imposing severe constraints in the allowed range of the primary hair values. We analyzed the null geodesics near this black hole, demonstrating how the scalar hair influences the shadow diameter. Specifically, the influence of primary hair on the shadow size is such that when the parameter of the Beyond Horndeski function $F_4$ is negative, an increase in scalar hair enlarges the shadow. However, when this parameter is positive, an increase in scalar hair reduces the shadow size. Additionally, we constrain the scalar hair using observational data, revealing its sensitivity to the choice of other black hole parameters. To further probe the observational features of the scalar hair, we simulate face-on two-dimensional images of spherically infalling accretion disks, examining how the primary scalar hair affects the black hole's shadow. Finally, we impose all relevant constraints to identify black holes that are both stable and consistent with observational data.
\bigskip

%{\noindent{\textit{Keywords}}: Beyond Horndeski theories, scalar hair, black hole thermodynamics, black hole shadow

%\noindent{PACS numbers}: 04.20.Fy, 04.20.Jb, 04.25.-g   
\end{abstract}

\maketitle
%%%%%%%%%%%%%%%%%%%%%%%%%%%%%%%%%%
\section{Introduction}

Black holes are among the most remarkable predictions of General Relativity (GR), and we are fortunate to live in an era when these predictions are being confirmed. Over the past decade, the existence of black holes has been substantiated through the development of advanced observational techniques and detection methods. The new era of gravitational-wave astronomy began with the first direct detection of gravitational waves and the first observation of a binary black hole merger~\cite{Abbott1}. The concurrent detection of gravitational waves and gamma rays from the collapse of a binary neutron star system further confirmed the luminal propagation of gravitational waves~\cite{Abbott2,Abbott3}. Moreover, the Event Horizon Telescope (EHT) provided the first direct images of black holes, specifically of the supermassive black holes (SMBHs) M87*~\cite{eht1,Akiyama:2019,the_event_horizon_telescope_collaboration_first_2019,the_event_horizon_telescope_collaboration_first_2019-1} and, a few years later, Sgr A*~\cite{Akiyama:2022}, located at the centers of galaxy M87 and our galaxy, respectively. The refinement of these observations has even revealed the polarization of the emission ring~\cite{eht3,ehtsa2}. These outstanding results have consolidated GR as the most successful theory of gravity for over a century.

However, GR is not a fundamental theory; it cannot provide satisfactory answers to some unresolved questions and fails to describe gravitational interactions at scales smaller than the Planck length or larger than astronomical units. On the one hand, we lack evidence on how gravitational interactions behave in the region between the Planck length and the micron scale, as gravitational experiments cannot access this domain. On the other hand, observations at larger length scales than those of the solar system pose challenges to GR. A concrete example is that GR only aligns with galactic, extragalactic, and cosmological data by requiring six times more dark matter than visible matter \cite{cc1,cc2}, to be consistent with the very small observed cosmological constant. In contrast, quantum field theory predicts a cosmological constant value 120 orders of magnitude larger, a significant discrepancy known as the cosmological constant problem~\cite{ccproblem}. Moreover, GR faces challenges at large scales, such as the Hubble tension, a four-sigma-level discrepancy between two independent measurements~\cite{tension2,tension1}. The first measurement is a model-independent experiment determining the recession velocities of type Ia supernovae~\cite{tension4,tension3}, while the second extrapolates data from the cosmic microwave background (CMB) using the Lambda cold dark matter model~\cite{tension5}.

There is hope that many of these issues will find resolution in a complete theory of quantum gravity, which remains elusive. The quest for such a fundamental theory of gravity has been long and complex. Nonetheless, several promising candidates have emerged, including String theory \cite{Zw}, emergent gravity \cite{verlinde}, asymptotic safety \cite{reuter}, and loop quantum gravity \cite{LQG}. A more consistent formulation of these theories could potentially resolve the problem of singularities \cite{singularity1,singularity2}, which are a robust prediction of GR under reasonable generic conditions of matter content \cite{Penrose}. Additionally, a complete understanding of the relationship between the laws of thermodynamics \cite{Bekenstein01,H1,H2} and gravity from a microscopic perspective could be achieved \cite{ashtekar,strominger}. Meanwhile, alternative theories of gravity can be viewed as modest attempts to address the current challenges in gravitational physics. Given the difficulty in obtaining experimental data relevant to quantum gravity, these alternative theories, thought of as effective field theories of an underlying fundamental theory, serve as phenomenological tools that bridge the gap between quantum gravity candidates and measurements at intermediate and large scales.

Modified gravity theories are systematic approaches developed in the low-energy limit of an as-yet-unknown UV-complete theory of gravity. This implies that when GR is considered an effective field theory, it incorporates corrections from additional fields. In this context, Horndeski theory was introduced as the most general scalar-tensor theory, leading to second-order field equations and being free of Ostrogradski instabilities~\cite{Horndeski:1974wa}. However, exhaustive research over the last decade has shown that Horndeski theory is not the ultimate framework. It was realized that scalar-tensor theories could generate higher-order field equations while still avoiding Ostrogradski instabilities by considering degenerate Lagrangians. This insight allowed for the generalization of Horndeski theory, first into Beyond Horndeski and later into Degenerate Higher Order Scalar Tensor (DHOST) theories, providing an enriched class of effective field theories~\cite{Crisostomi:2016tcp,Gleyzes:2014dya,Kobayashi:2019hrl,Langlois:2015cwa,Langlois:2017mdk,Langlois:2018dxi}.

Finding black hole solutions with a nontrivial scalar field in these theories is a complex task, primarily due to no-hair extensions in scalar-tensor theories that prevent the existence of nontrivial local solutions~\cite{SotiriouBHs,Hui,MaselliBHs}. The no-hair conjecture suggests that the final state of gravitational collapse, in the presence of any matter-energy, is a black hole characterized solely by its mass $M$, angular momentum $J$, and electric charge $Q$, quantities that follow a Gauss law and are measured at infinity. When two black holes share the same $M$, $J$, and $Q$, they are described by the same Kerr-Newman metric~\cite{PhysRevLett.11.237}. These black holes are said to have ``no hair", meaning they possess no other independent, conserved charge~\cite{Bekenstein:1971hc,Teitelboim:1972ps,Chrusciel:2012jk,Mazur:2000pn}. If a black hole is characterized by an additional global charge distinct from mass, angular momentum, or electric charge, it possesses primary hair. A black hole exhibits secondary hair when the metric is entirely determined by $M$, $J$, and $Q$, even though the configuration contains nontrivial additional fields (other than electromagnetic). These configurations are often called ``hairy black holes'', even when they possess secondary hair~\cite{Herdeiro:2015waa}.

Constructing black hole solutions with primary hair is considerably more challenging than those with secondary hair. Only recently, in ~\cite{Bakopoulos:2024}, the first black hole solution with primary hair was obtained in the shift-symmetric subclass of Beyond Horndeski theories, with a subsequent generalization \cite{Baake:2023zsq,Bakopoulos:2023sdm}. The primary hair is the conserved scalar charge associated with the shift symmetry of the theory. This symmetry allows the scalar field to have different symmetries from the static black hole's spacetime, facilitating the derivation of an analytical solution~\cite{Babichev:2013cya}. Given its recent discovery, the physical implications of this black hole with primary hair are still being explored. Thus far, research has been limited to the thermodynamic properties of certain special cases in its generalization~\cite{Bakopoulos:2024ogt} and the description of its weak gravitational lensing~\cite{Mushtaq:2024qse}. However, investigating the shadow of these black hole configurations remains unexplored, providing an excellent opportunity to advance our understanding of the primary hair. Furthermore, from a thermodynamic perspective, it can be interesting to analyze to what extent primary hair plays a role in black hole stability. This motivates the main goal of this research, which is focused on studying the first black hole with primary hair in the shift symmetric subclass of Beyond Horndeski theories found in~\cite{Bakopoulos:2024}.

For these purposes, this work covers three aspects. Following the order of this analysis: First, to identify sectors in the space of parameters where locally stable black holes are possible from a thermodynamic perspective. Second, to establish the first constraints on the theory's parameters where black hole solutions are compatible with the M87* and Sgr A* shadows observed by the EHT collaboration. Third, to investigate how the primary hair affects other observational signatures, such as the accretion disk.

Observations of the shadows cast by SMBHs have provided an unprecedented opportunity to explore modified theories of gravity in the strong-field regime. This opens up the possibility of describing astrophysical black holes with alternatives to the Kerr metric, considering that astronomical observations suggest nearly 70\% of all stellar black holes are near-extremal~\cite{Aretakis:2018dzy}. However, for simplicity, throughout this work, we will neglect the effect of rotation and focus on the static, spherically symmetric, and asymptotically flat black hole solution found in \cite{Bakopoulos:2024}. The justification for this choice is that the effect of rotation on the shadow radius is negligible at all inclination angles. For Kerr black holes, the shadow's size is marginally smaller than a Schwarzschild black hole (SBH) and depends primarily on mass $M$, with minor contributions from spin and inclination angle. While the effect is more pronounced at higher inclination angles, where the shadow becomes slightly asymmetric along the spin axis, even in the edge-on view of the extremal case, the difference of the shadow size with respect to Schwarzschild remains small, around 12\%~\cite{vagnozzi_horizon-scale_2023}. The spin influences the shadow's circularity, but there is still no consensus on its precise impact~\cite{Vagnozzi:2022tba}. Based on these considerations, it is reasonable to assume that the effects of rotation also remain small in our case. However, this must be confirmed once a rotating version of this black hole is available.

The paper is organized as follows. In Sec. \ref{sec:BH_solution}, we review the black hole solution in the context of shift-symmetric Beyond Horndeski gravity. In Sec. \ref{sec:thermo}, we explore the thermodynamic properties of the black hole, including temperature profiles, entropy, and heat capacity, when the scalar hair and the theory's parameters vary. In Sec. \ref{sec:null}, we study light propagation around the black hole using a standard Lagrangian formalism, calculate the theoretical shadow diameter, and demonstrate how primary scalar hair can be constrained using EHT observational data. The sensitivity of the shadow diameter to changes in other black hole parameters is also examined. In Sec. \ref{sec:infall}, we classify the photon orbits around the black hole and explain how different orbits result in varying images of a luminous source. We then simulate face-on images of an accretion disk around the black hole using ray-tracing methods, classify the light rings that confine the shadow, and investigate the observational features of infalling spherical accretion. Finally, we summarize our findings in Sec. \ref{sec:conclusions}.

Throughout this study, we use geometric units with $8\pi G=\hbar=\ell_p=c=1$, adopt the sign convention $(-,+,+,+)$, and use primes to denote derivatives with respect to the radial coordinate.
%%%%%%%%%%%%%%%%%%%%%%%%%%%%%%%%%%%sect.1
\section{Brief review on the black hole solution with primary scalar hair}\label{sec:BH_solution}

{This section provides an overview of the key features of the static, spherically symmetric black hole solution found in Ref.~\cite{Bakopoulos:2024}.} In the framework of Beyond Horndeski theories \cite{Gleyzes:2015}, where the additional shift symmetry $\varphi\rightarrow\varphi+\mathrm{const.}$ and the parity symmetry $\varphi\rightarrow{-\varphi}$ are present, the theories are parametrized by three arbitrary functions $G_2$, $G_4$ and $F_4$, which are given in terms of the kinetic term of the scalar field  $X=-\frac{1}{2}\varphi_{,\mu}\varphi^{,\mu}$. This action is given by
\begin{equation}
I\left[g_{\mu\nu},\varphi\right]=\frac{1}{2%\kappa
}\int\ed^4x\,\sqrt{-g}\Biggl\{
G_2(X)+G_4(X)R+G_{4 X}\left[(\square \varphi)^2-\varphi_{\mu \nu} \varphi^{\mu \nu}\right]+F_4(X) \epsilon^{\mu \nu \rho \sigma} \epsilon^{\alpha \beta \gamma}{ }_\sigma \varphi_\mu \varphi_\alpha \varphi_{\nu \beta} \varphi_{\rho \gamma}
\Biggr\},
    \label{eq:action}
\end{equation}
where the following notations are adopted for conciseness: $\varphi_\mu=\partial_\mu \varphi$, $\varphi_{\mu \nu}=\nabla_\mu \partial_\nu \varphi$ and derivation with respect to $X$ is represented by a subscript $X$. The Horndeski functionals that support the hairy black hole are given by,
\begin{equation}
G_2=-\frac{8\eta}{3\lambda^2}X^2,\qquad G_4=1-\frac{4\eta}{3}X^2,\qquad F_4=\eta\ ,
    \label{eq:functionals}
\end{equation}
parametrized by $\lambda$, with dimension of length and $\eta$, with dimension (length)$^4$. The scalar field has the functional form,
\begin{equation}
\varphi = q t +\psi(r),
    \label{eq:varphi}
\end{equation}
in which $q$ has dimension (length)$^{-1}$ and is identified as the primary hair. The static spherically symmetric spacetime that describes the black hole is given by the line element
\begin{equation}
\ed s^2=-f(r)\ed t^2+\frac{\ed r^2}{f(r)}+r^2\ed\theta^2+r^2\sin^2\theta\ed\phi^2,
    \label{eq:metr0}
\end{equation}
whose metric function is given by
\begin{equation}
f(r) = 1-\frac{2M}{r}+\mc\left[\frac{\pi/2-\arctan(r/\lambda)}{r/\lambda}+\frac{1}{1+(r/\lambda)^2}\right]\ ,
    \label{eq:lapse_0}
\end{equation}
where $\mathcal{c}=\eta q^4$ is a dimensionless parameter. The expression of $\psi(r)$ is determined by the differential equation
\begin{equation}
\psi'(r)=\pm\sqrt{\frac{q^2}{f(r)^2}\left[1-\frac{f(r)}{1+(r/\lambda)^2}\right]}\ ,
    \label{eq:psi}
\end{equation}
with prime indicating differentiation with respect to the radial coordinate. Thus, the kinetic term of the scalar field is
\begin{equation}
X=\frac{q^2/2}{1+(r/\lambda)^2}\ .
    \label{eq:X}
\end{equation}
Since the theory is invariant under the transformation $\lambda\rightarrow-\lambda$, we can fix $\lambda$ to be positive. 
Note that, for $M=\mc=0$, the solution reduces to the flat spacetime, while for $M=0$ and $\mc\neq0$, we obtain a nontrivial zero-mass spacetime which for $\mc<0$ describes a black hole, and for $\mc>0$ describes a naked singularity. 

In the metric function \eqref{eq:lapse_0}, the two integration constants $M$ and $q$, correspond, respectively, to the Arnowitt-Deser-Misner (ADM) mass \cite{ADM:1961}, and the primary scalar hair. Looking at Eq. \eqref{eq:X}, we see that when $q=0$ (indicating the absence of scalar hair), the modifications to GR disappear, and the solution in Eq. \eqref{eq:lapse_0} reverts to the Schwarzschild solution.

%The asymptotic behavior of the metric function as $r\rightarrow\infty$ is
At large distances, the lapse function \eqref{eq:lapse_0} can be estimated as
\begin{equation}
f(r) = 1-\frac{2M}{r}+2\lambda^2\frac{\mc}{r^2}+\mathcal{O}(r^{-4}),
    \label{eq:lapse_1}
\end{equation}
which correspond to the asymptotic scalar fields $\varphi=q v$ and $\varphi=q u$, respectively, for the adoption of $+$ and $-$ signs in Eq. \eqref{eq:psi}, where $u$ and $v$ are the advanced and retarded null coordinates. The solution \eqref{eq:lapse_1}, considering the positive ADM mass $M$, is asymptotically flat and resembles the RN solution in GR, with $\mc$ playing the role of the black hole's electric charge. 

\begin{figure}[t]
    \begin{subfigure}[t]{0.45\textwidth}
    \includegraphics[width=7cm]{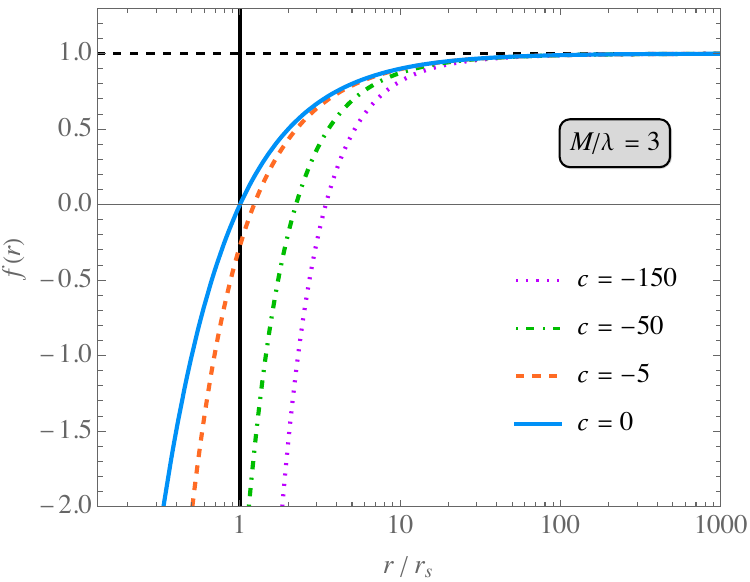}
    \captionsetup{justification=centering}
    \caption{}
    \label{fig:f(r)a}
    \end{subfigure}
    \begin{subfigure}[t]{0.45\textwidth}
    \includegraphics[width=7cm]{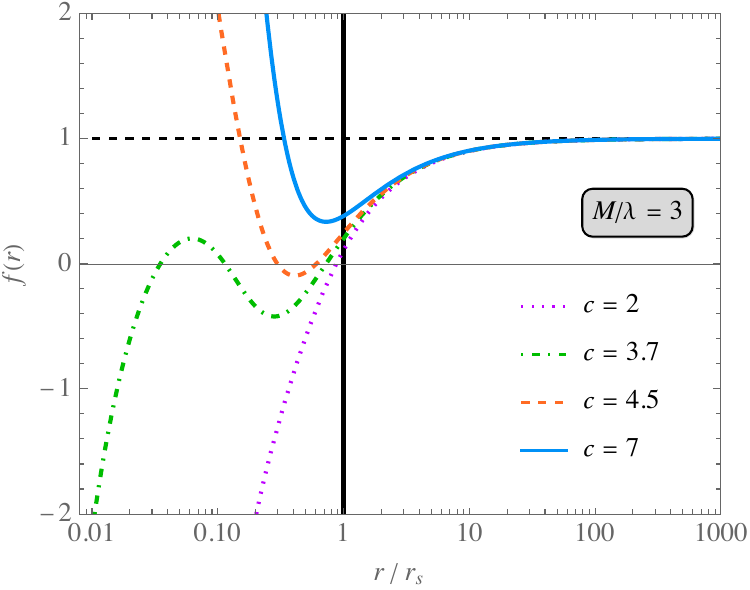}
    \captionsetup{justification=centering}
    \caption{}
    \label{fig:f(r)b}
    \end{subfigure}
    \caption{The behavior of the lapse function $f(r)$ for $\mc\leq0$ in panel (a), and $\mc>0$ in panel (b), plotted various values of $\mc$. While the case of $\mc<0$ corresponds to a single-horizon black hole, the case of $\mc>0$ demonstrates various possibilities ranging from multiple-horizon black holes to naked singularities. The horizontal axis is logarithmic in both panels, and the unit of length along the axes is set as the Schwarzschild radius $r_s$.}
    \label{fig:f(r)}
\end{figure}

The sign of the coefficient $\mc$ is determined by the sign of $\eta$, which is the most important characteristic of this latter parameter since to have a fixed $\mc$, every change in $\eta$ has to be accounted for by the scalar hair $q$. Note also that the sign of $\eta$ is also crucial in the identification of the causal structure of the black hole, as the number of horizons depends strictly on this parameter. In fact, in the limit $r\rightarrow0$, the lapse function \eqref{eq:lapse_0} is reduced to \cite{Bakopoulos:2024}
\begin{equation}
f(r) = 1-\frac{2M-\pi\mc\lambda/2}{r}-\frac{2\mc r^2}{3\lambda^2}+\mathcal{O}(r^4).
    \label{eq:lapse_2}
\end{equation}
In Fig. \ref{fig:f(r)} we have plotted the radial profile of the lapse function \eqref{eq:lapse_0}, for different values of the coefficient $\mc$, and considering the both cases of negative and positive $\mc$.
Schwarzschild black hole appears when $\mc=0$ and as reference $r_s=2M$ is the Schwarzschild radius. As inferred from Fig. \ref{fig:f(r)a}, for $\mc<0$, $f(r)\rightarrow-\infty$ as we approach the black hole's singularity at $r=0$. In this scenario, the black hole has one event horizon and $r_+>r_s$, meaning that black holes with negative primary hair are larger than Schwarzschild black holes. For $\mc>0$, the theory can propose different scenarios, ranging from black holes with multiple horizons to naked singularities. As observed from Eq. \eqref{eq:lapse_2}, and accordingly in Fig. \ref{fig:f(r)b}, this fact strictly depends on the value of the ratio $M/\lambda$ compared to the ratio $\pi\mc/4$ determining a critical value of $\mc$ given by $\mc_*=4 M/(\pi\lambda)$. If $\mc<\mc_*$, we observe $f(r)\rightarrow-\infty$ as we approach the singularity. In this case, the event horizon $r_+$ is inside $r_s$; hence, the black hole is more compact than its GR counterpart with $\mc=0$. Increasing the value of $\mc$, the event horizon shrinks until three horizons appear provided $M/\lambda>1+\pi/4$ \cite{Bakopoulos:2024}. When the primary hair is sufficiently large, namely $\mc>\mc_*$, $f(r)\rightarrow+\infty$ as we approach the singularity. As $\mc$ increases, the condition becomes less favorable for having the horizons. First, we obtain a black hole with two horizons, followed by an extremal black hole (EBH), and finally, for larger $\mc$, a naked singularity appears.

As a final remark in this section, we note that the theory is defined by $\lambda$ and $\eta$, and, as it is clear from metric function in Eq. \eqref{eq:lapse_0}, it is this dimensionless parameter $\mc$, along with $M/\lambda$ the relevant quantities that we use in the analyses carried out in the following sections. These sections cover the thermodynamic aspects and the observational signatures of these black holes.

%%%%%%%%%%%%%%%%%%%%%%%%sect.2
\section{Local thermodynamic stability}\label{sec:thermo}

The black hole solution under study possesses two theory parameters, $\lambda$ and $\eta$, and two integration constants, $c$ and $M$. In this section, the consequences role of the primary hair in the local thermodynamic stability is performed in two ways: by fixing $M/\lambda$ and by fixing $\mc=\eta q^4$.

\begin{figure}[h!]
    \centering
    \includegraphics[width=7cm]{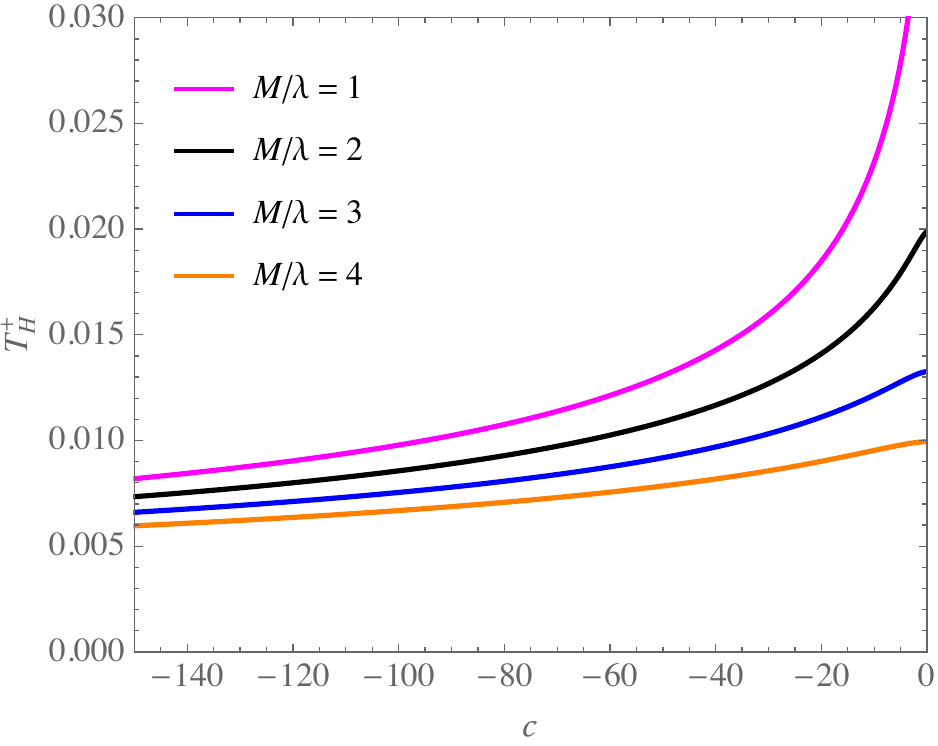} (a)\qquad
    \includegraphics[width=7cm]{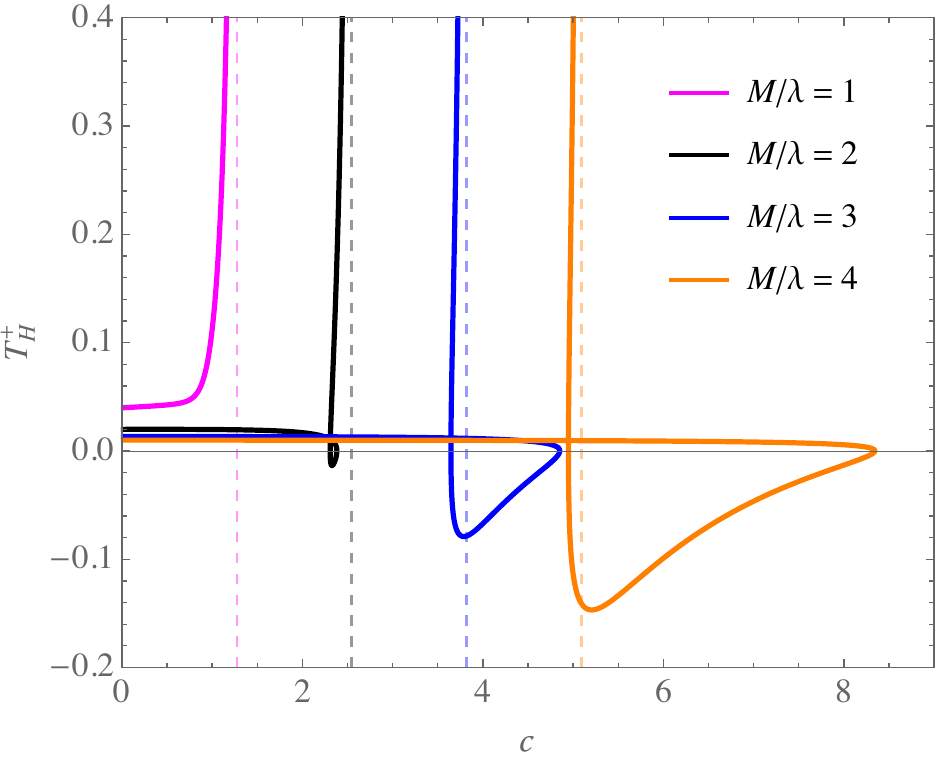} (b)
    \caption{The $\mc$-profile of the Hawking temperature $T_{\rm{H}}^+$ plotted for (a) $\mc<0$ and (b) $\mc>0$, with four different values of the ratio $M/\lambda$. For the SBH case with $M/\lambda=1$, the corresponding temperature is $T_{\rm{H}}^+=1/8\pi\approx0.393$, as expected. In panel (b), the $\mc$ values corresponding to the asymptotic curve where temperatures tend to infinity are $\mc_{*}=\{1.273, 2.546, 3.819,5.092\}$, for $M/\lambda=\{1,2,3,4\}$, respectively.}
    \label{fig:TH}
\end{figure}

It is well-known that the Hawking temperature of a static spherically symmetric black hole spacetime defined by the line element \eqref{eq:metr0}, is given by $T_{\mathrm{H}}^+=\kappa_+/(2\pi)$ \cite{Hawking:1974sw}, in which $\kappa_+=f'(r_+)/2$ is the surface gravity of the black hole on its event horizon \cite{nielsen_dynamical_2008,Pielahn:2011}. Hence, by using the lapse function \eqref{eq:lapse_0}, we get
\begin{equation}\label{eq:TH_1}
    T_{\rm{H}}^+(r_+)=\frac{r_+^4-2(\mc-1) r_+^2 \lambda^2+\lambda^4}{4 \pi r_+\left(r_+^2+\lambda^2\right)^2}.
\end{equation}
Since $r_+$ is a function of $M$, $\lambda$, and $\mc$, the behavior of the Hawking temperature with respect to the primary hair $\mc$ can be readily seen by fixing the $M/\lambda$ ratio. Different cases have been plotted in Fig. \ref{fig:TH}. 
%
%%% description of TH+, regarding its strange behavior
One can observe that when $\mc<0$ (Fig. \ref{fig:TH}(a)), the Hawking temperature rises smoothly as $\mc$ tends to zero. It is clear that when $\mc=0$, the temperature correctly reduces to $T_{\rm{H}}^+=1/(4\pi r_s)$, which corresponds to the SBH's temperature. This indicates that the SBH possesses the highest Hawking temperature for negative primary hair. We also notice that the temperature decreases as the ratio $M/\lambda$ increases, resulting in a colder black hole. The behavior of the temperature profile gets more involved for $\mc>0$, where the temperature reaches a local maximum and then a local minimum for sufficiently large values of the $M/\lambda$ ratio before abruptly reaching a large temperature at an asymptotic value $\mc_{*}$ for the primary hair which is where the temperature tends to infinite. This asymptotic value for $\mc$ is mapped to $r_+\rightarrow0$ because the $M/\lambda$ ratio is fixed. This can be extracted from the $f(r_+)=0$ condition when $r_+\rightarrow0$, given $\mc_{*}=4 M/(\pi\lambda)$. The minimum temperature can be lower than zero, discarding one branch of solutions between the extremal configurations as they are not physical. For a sufficiently small value of the $M/\lambda$ ratio, all sizes of the black holes get positive temperatures. There is one critical value for the $M/\lambda$ ratio with only one EBH.

\begin{figure}[h!]
    \begin{subfigure}[t]{0.45\textwidth}
    \includegraphics[width=7cm]{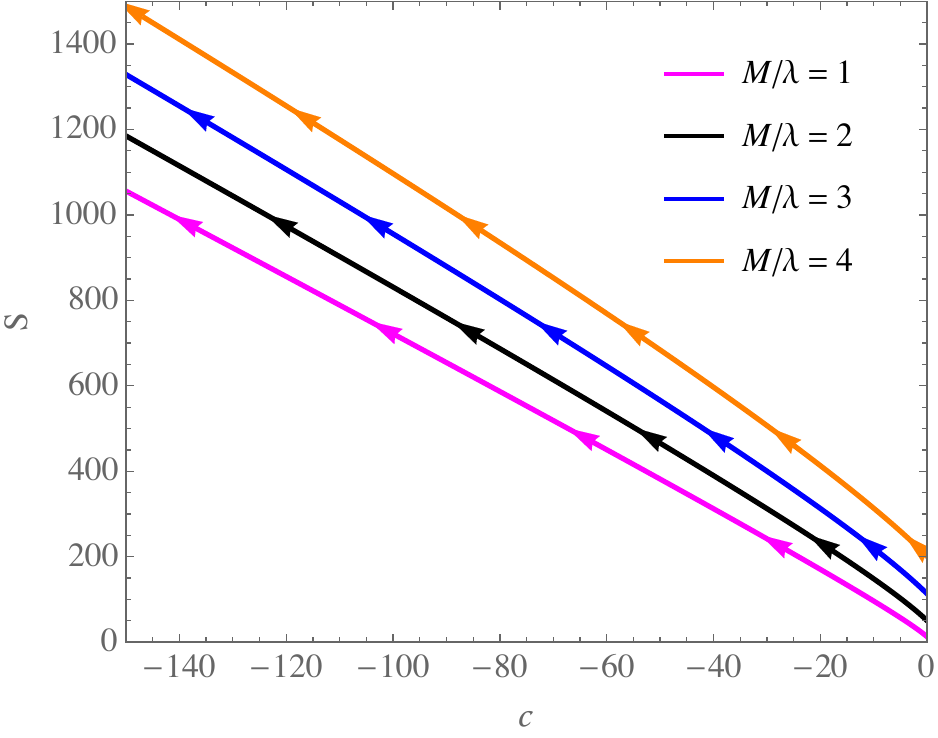}
    \captionsetup{justification=centering}
    \caption{}
    \label{svsc1}
    \end{subfigure}
    \begin{subfigure}[t]{0.45\textwidth}
    \includegraphics[width=7cm]{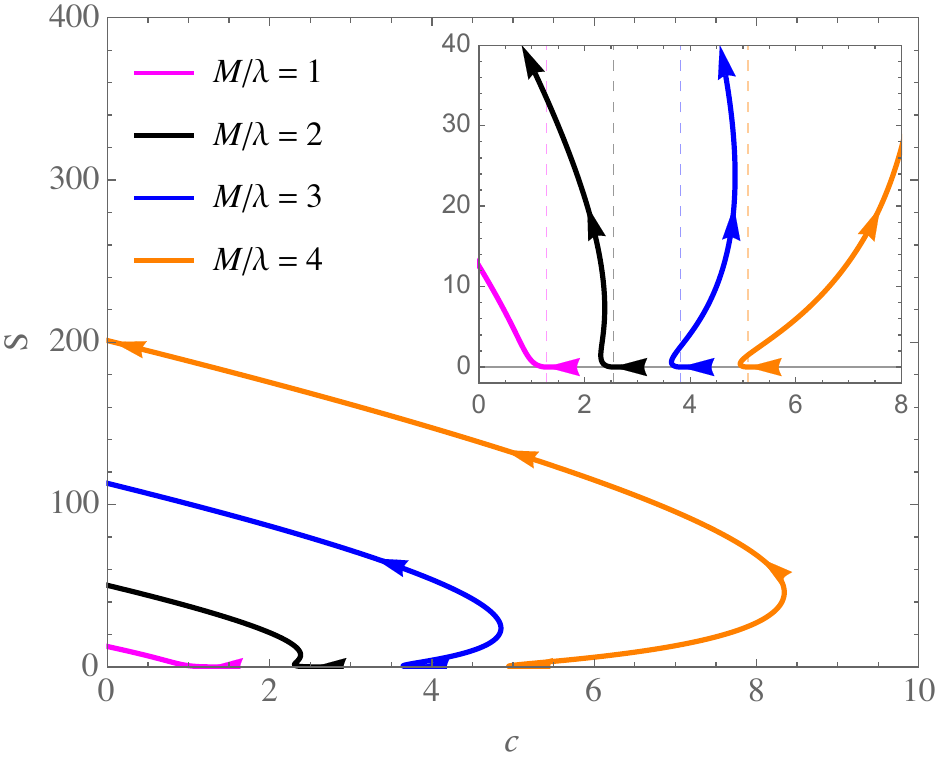}
    \captionsetup{justification=centering}
    \caption{}
    \label{svsc2}
    \end{subfigure}
    \caption{The Bekenstein-Hawking entropy respect to $\mc$ plotted for (a) $\mc<0$ and (b) $\mc>0$, with four different values of the ratio $M/\lambda$. The arrows show the direction in which $r_+$ increases. The entropy is a multivalued function for sufficiently large and positive values of $\mc$.}
    \label{fig:B-H_S}
\end{figure}

Note that, in our adopted system of units, the Bekenstein-Hawking entropy for a spherically symmetric black hole is given by the area law \cite{bekenstein_bekenstein-hawking_2008}
\begin{equation}
S = \pi r_+^2,
    \label{eq:B-H_S}
\end{equation}
whose $\mc$-profile has been plotted in Fig. \ref{fig:B-H_S}, for the both cases of $\mc>0$ and $\mc<0$. 
%
%%% description of S, regarding its strange behavior
The diagrams show that the entropy monotonically decreases in the negative $\mc$ domain (cf. Fig. \ref{svsc1}). However, as it is shown in Fig. \ref{svsc2}, for positive values of $\mc$ and sufficiently large values of the $M/\lambda$ ratio, the entropy gets multivalued for tiny black holes, changing its slope. This is in the neighborhood close to the asymptotic values $\mc_*$ described in Fig. \ref{fig:TH}(b). This reflects the fact that as $\mc$ approaches $\mc_*$, the black hole size $r_+$ tends to zero, and consequently, the entropy given by \eqref{eq:B-H_S} also tends to zero. The arrows in these plots represent the direction along which $r_+$ increases. In general, for any value of the primary hair, it is observed that the higher the ratio $M/\lambda$, the more entropic the black hole becomes.

\begin{figure}[h!]
    \centering
    \begin{subfigure}[t]{0.5\textwidth}
        \centering
        \includegraphics[width=7cm]{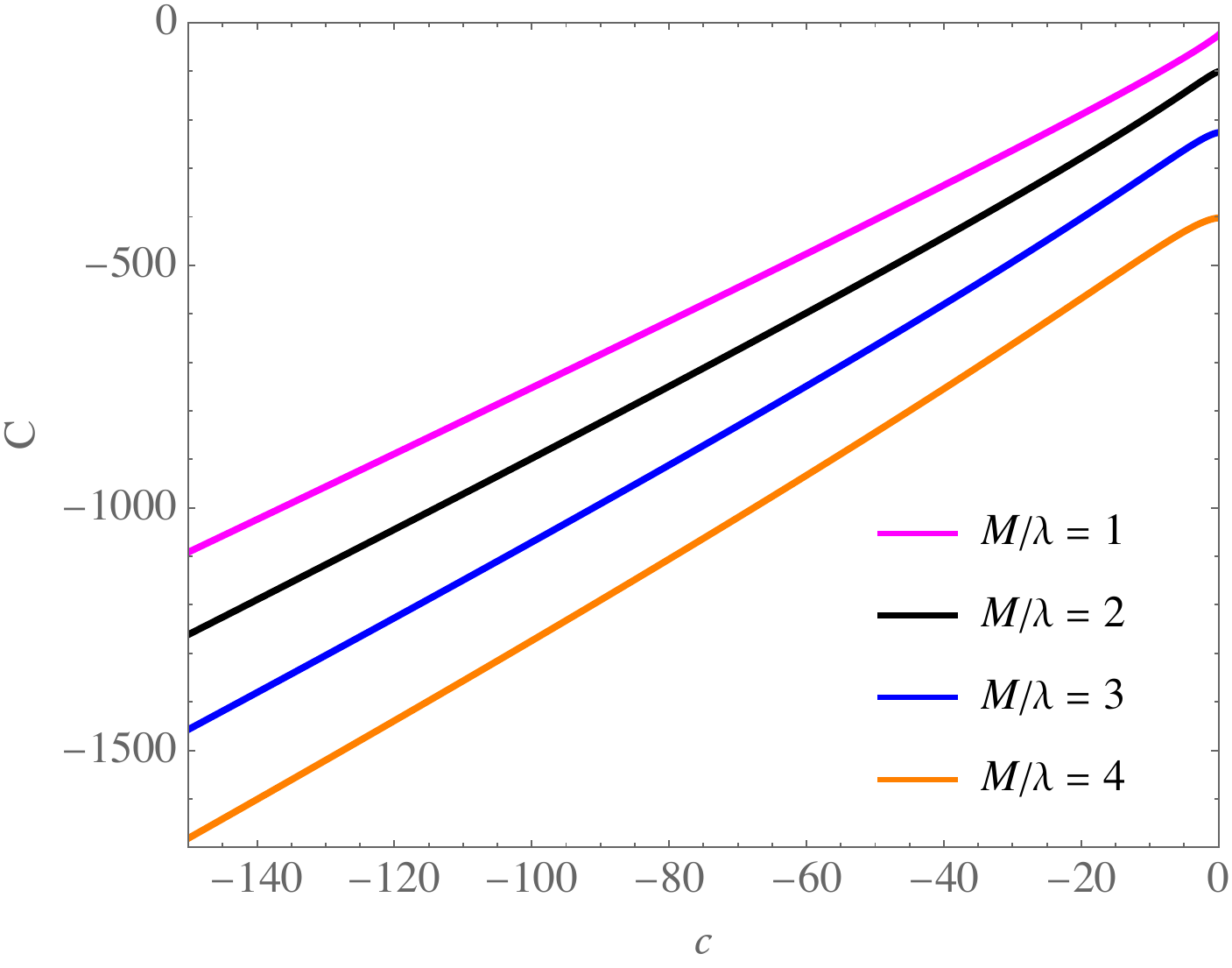}
        \captionsetup{justification=centering}
        \caption{}
        \label{hcvsca}
    \end{subfigure}%
    ~ 
    \begin{subfigure}[t]{0.5\textwidth}
        \centering
        \includegraphics[width=7cm]{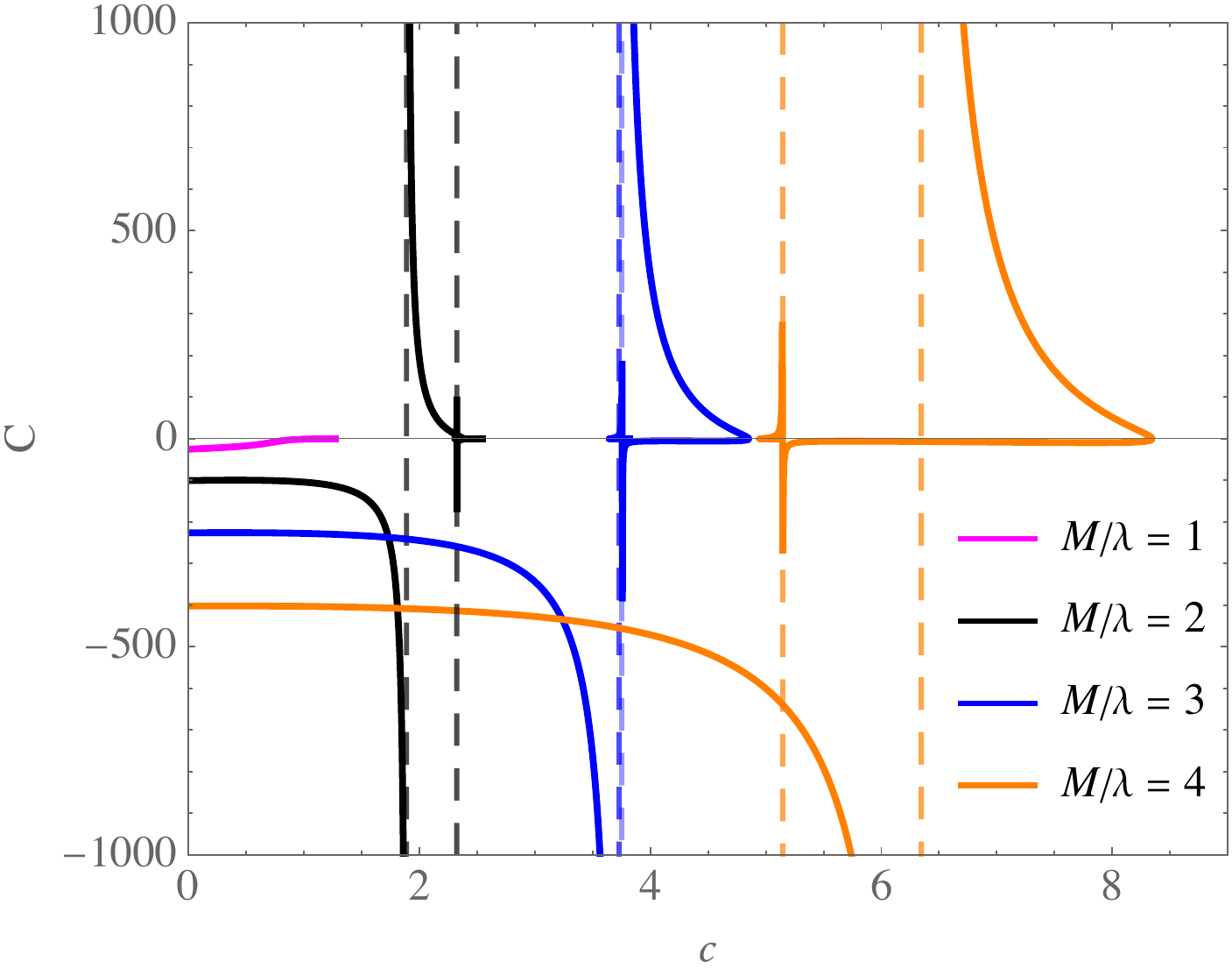}
        \captionsetup{justification=centering}
        \caption{}
        \label{hcvscb}
    \end{subfigure}

    \begin{subfigure}[t]{0.7\textwidth}
        \centering
        \includegraphics[width=7cm]{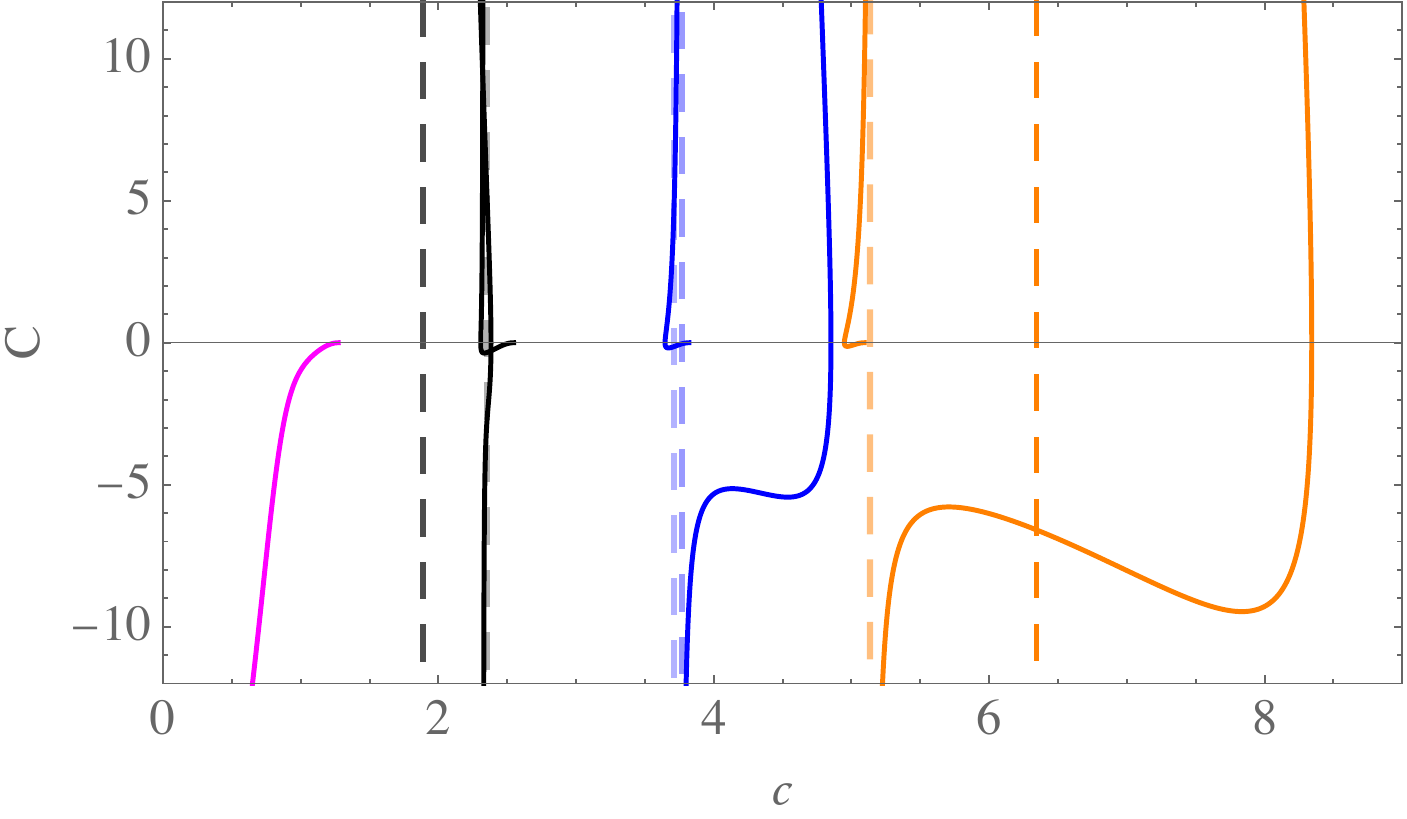}
        \captionsetup{justification=centering}
        \caption{}
        \label{hcvscc}
    \end{subfigure}
    \caption{Heat capacity as a function of $\mc$ for different values of $M/\lambda$. The negative domain of $\mc$ (Fig. \ref{hcvsca}) shows negative heat capacity. The positive domain of $\mc$ exhibits branches with positive heat capacity in Figs. \ref{hcvscb} and \ref{hcvscc}.}
    \label{hcvsc}
\end{figure}

The local stability of a black hole can be analyzed by studying its response under small perturbations of its thermodynamic variables around equilibrium. For black holes with primary hair, it is convenient to analyze the heat capacity. It is given by
\begin{equation}
C=T_{\rm{H}}^+\frac{\partial S}{\partial T_{\rm{H}}^+}=T_{\rm{H}}^+\left(\frac{\partial S}{\partial r_+}\right)\left(\frac{\partial T_{\rm{H}}^+}{\partial r_+}\right)^{-1}, 
    \label{eq:heatCap_0}
\end{equation}
which using Eqs. \eqref{eq:TH_1} and \eqref{eq:B-H_S} provides
\begin{equation}
C=-\frac{2 \pi r_+^2\left(r_+^2+\lambda^2\right)\Bigl(r_+^4-2(\mc-1) r_+^2 \lambda^2+\lambda^4\Bigr)}{r_+^6+3(1-2\mc) r_+^4 \lambda^2+(3+2\mc) r_+^2 \lambda^4+\lambda^6}.
    \label{eq:heatCap_1}
\end{equation}
It is straightforward to check that 
\begin{equation}
\lim_{\mc\rightarrow0 \atop} C=-2\pi r_+^2,
    \label{eq:Cp_Sch}
\end{equation}
which is the value corresponding to the SBH with $r_+=r_s$. In Fig. \ref{hcvsc}, the heat capacity profiles have been plotted within both the negative and positive domains of the $\mc$-parameter.
From \eqref{eq:heatCap_0}, it is clear that the sign of the heat capacity is given by the sign of $\partial T_{\rm{H}}^+/\partial r_+$ as the entropy is a monotonically increasing function of $r_+$. This fact can be clearly seen in agreement with Fig. \ref{hcvsc}, in the curves of $T_H^+$ respect to $r_+$ for fixed $M/\lambda$ radius. Namely,
\begin{figure}[h!]
    \centering
    \includegraphics[width=7cm]{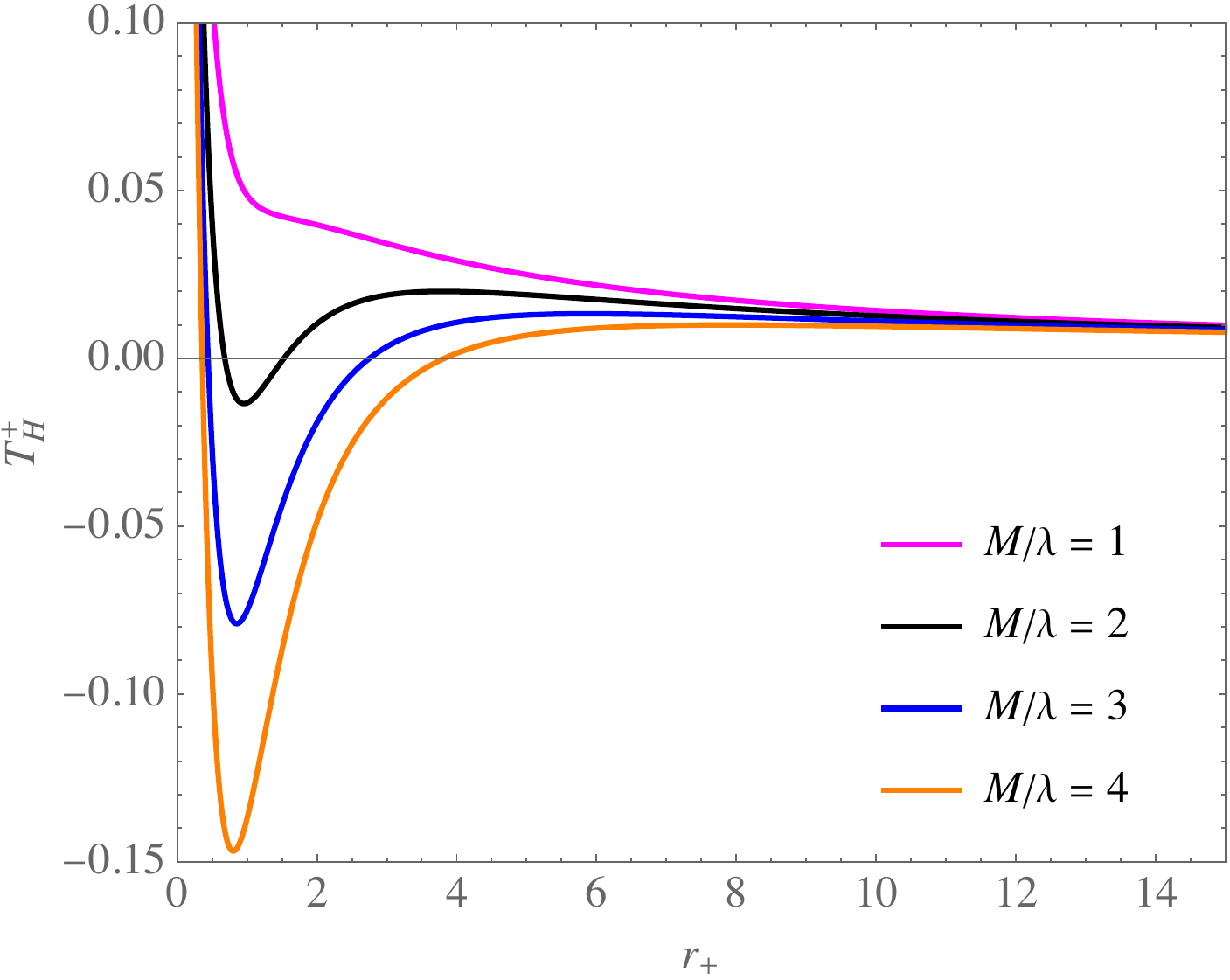}
    \captionsetup{justification=centering}
    \caption{Temperature profile as a function of $r_+$ for fixed values of the $M/\lambda$ ratio.}
    \label{fig:tvsrh1}
\end{figure}
Consequently, black holes with negative primary hair have negative heat capacity when $M/\lambda$ is fixed, as shown in Fig. \ref{hcvsca}. These are the large-sized black holes at the tail of the curves in Fig. \ref{fig:tvsrh1} after the temperature reaches a local maximum. However, in the positive sector of $\mc$, this is not the case for sufficiently large values of $M/\lambda$, where two branches with positive heat capacity emerge around two asymptotes where heat capacity diverges (cf. Fig. \ref{hcvscb}). We see this in more detail in Fig. \ref{hcvscc}, where the ratios $M/\lambda=\{2,3,4\}$ exhibit divergences in the heat capacity determined by the local maximum and minimum present in the temperature profiles (cf. Fig. \ref{fig:tvsrh1}). Of these two branches with positive heat capacity, only one is physical, as one possesses a negative temperature, corresponding to the branch departing from the minimum temperature with a negative slope. Therefore, locally stable black holes can be obtained for positive values of the primary hair and sufficiently large values of the $M/\lambda$ ratio. In Fig. \ref{hcvscb} and \ref{hcvscc}, this branch goes from the largest EBH to the vertical asymptote where the heat capacity diverges. This stable branch can be identified in the temperature curve as the section that departs from the largest EBH until the black hole with the local maximum temperature. For values of $M/\lambda<1+\pi/4$, there are no EBHs, and this locally stable branch starts from the minimum temperature until the maximum temperature. However, if the $M/\lambda$ ratio is too small, branches with positive heat capacity disappear as in the case $M/\lambda=1$.  When the ratio $M/\lambda=1+\pi/4$, the first EBH emerges with a vanishing minimum temperature. We can conclude that the higher the $M/\lambda$ ratio, the larger the branch of locally stable black holes.

\begin{figure}[h!]
   \centering
 \includegraphics[width=0.5\linewidth]{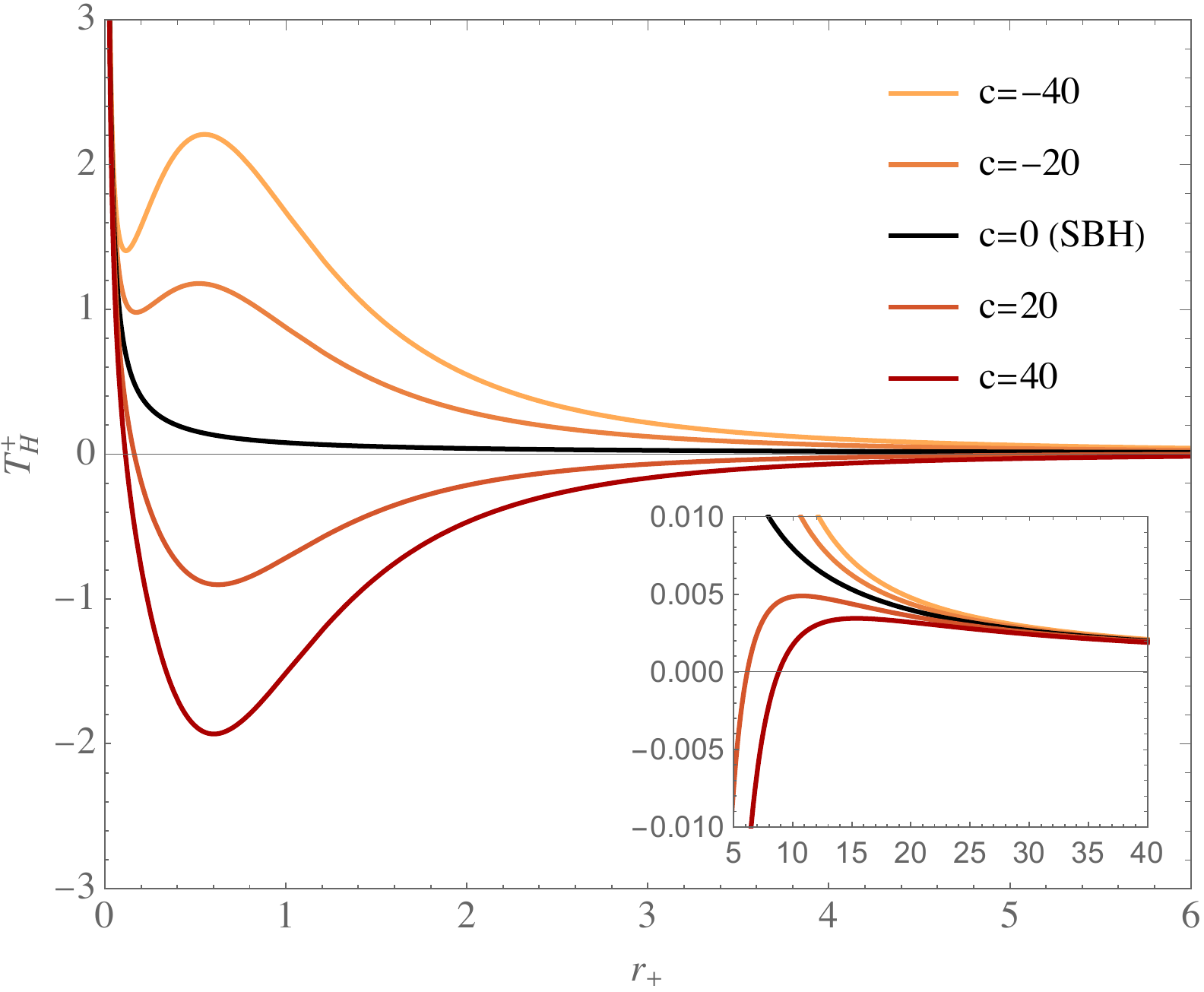}
    \caption{Temperature in terms of the black hole size $r_+$ and different values of the primary hair. The black curve represents the SBH temperature profile with vanishing primary hair. In these curves, $\lambda=1$.}
    \label{fig:tvsrh}
\end{figure}

The previous analysis has considered a fixed $M/\lambda$ ratio, and it is qualitatively equivalent to negative values of $\lambda$, since Eqs. \eqref{eq:TH_1} and \eqref{eq:heatCap_1} are parity even functions of $\lambda$. If this ratio is not fixed and $\mc$ is fixed instead, taking into account expression \eqref{eq:TH_1}, the temperature profiles are given by Fig. \ref{fig:tvsrh}.
As a reference, the SBH is depicted in black. For any value of the primary hair, the negative slope in the temperature profile indicates negative heat capacity for tiny ($r_+\ll0$) and very large black holes ($r_+\gg 0$). This is because the temperature profile asymptotically resembles the SBH in such sectors. The primary hair plays an important role at intermediate sizes, as black holes with sufficiently negative primary hair develop an intermediate branch with positive specific heat. However, as we increase the primary hair, this stable branch disappears in a range of only unstable black holes that contain SBH, as expected. A unique EBH with $T_{\rm{H}}^+ = 0$ emerges when the scalar hair is $\mc = 2$ and corresponds to a triple horizon, located at $r_+=\lambda$ with $M/\lambda=1+\pi/4$, as described in \cite{Bakopoulos:2024}. As $\mc$ gradually increases, two EBHs are possible. The intermediate branches between these EBHs are non-physical as they acquire negative temperature, leaving the tiny and large branches disconnected and bounded by the EBHs. For sufficiently large positive values of the primary hair, a stable branch reappears, but only very large black holes are physical, as they have positive temperatures. This branch goes from the largest EBH to the black hole with the local maximum temperature (see the magnified plot in Fig. \ref{fig:tvsrh}), decaying into an unstable branch as the size increases and approaches SBH asymptotic. Namely, there is a range for primary hair where there are no locally stable black holes and is given by
\begin{equation}
    -\frac{1}{9}(13 \sqrt{13}+35)<\mc<\frac{1}{9}(13 \sqrt{13}-35)\ .
\end{equation}
This means locally stable black holes exist only for $\mc\lessapprox-9.097$ and $\mc\gtrapprox1.319$. This bound is found by analyzing the condition $T'(r_+)=0$. Since stable branches exist between the minimum and maximum temperatures, it is possible to analyze a cubic polynomial on $r_+^2$ that results from this condition and determine the critical values where the two positive roots of the polynomial are double.

%%%%%%%%%%%%%%%%%%%
\section{Null geodesics around a black hole with primary scalar hair and the shadow constraints}\label{sec:null}

\begin{figure}[t]
    \centering
    \includegraphics[width=7cm]{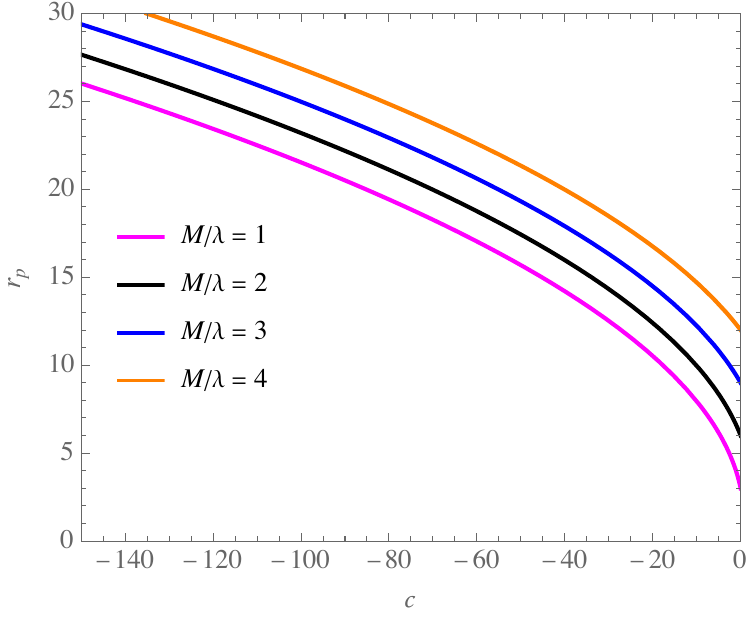} (a)\qquad
    \includegraphics[width=7cm]{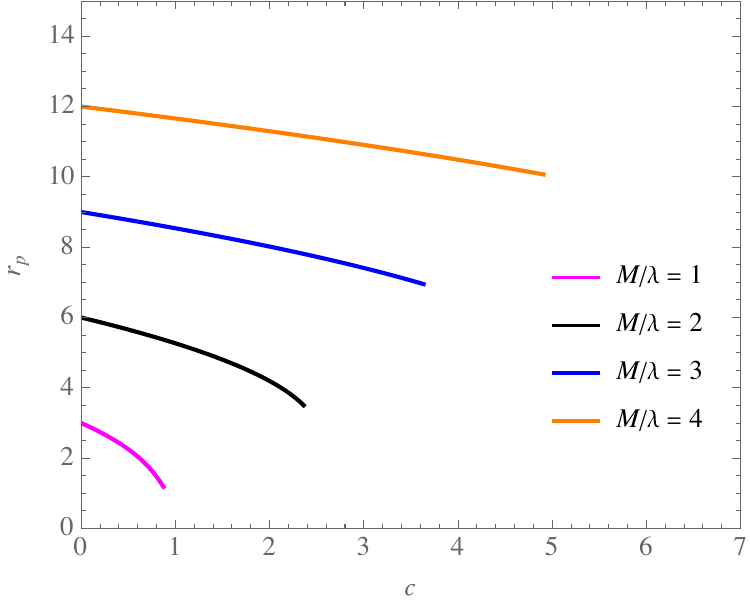} (b)
    \caption{ The $\mc$-profile of the photon sphere radius $r_p$, plotted for (a) $\mc<0$, and (b) $\mc>0$, and different values of the ratio $M/\lambda$.}
    \label{fig:rp}
\end{figure}

In this section, we obtain the boundary of the black hole shadow observable by a distant viewer and investigate the impact of the scalar hair on its diameter. In fact, the black hole shadow is defined by the photon region, distinguishing between captured and deflected light orbits. For a static spherically symmetric spacetime, this region corresponds to the photon sphere. Further in this section, we use the EHT data for M87* and Sgr A*, and constrain the $\mc$ and $\lambda$-parameters. To proceed with this task, let us define the Lagrangian 
\begin{equation}
\mL(\bm{x}, \dot{\bm{x}}) = \frac{1}{2} g_{\mu\nu} \dot{x}^\mu \dot{x}^\nu,
    \label{eq:Lagrangian}
\end{equation}
in which $\dot{\bm{x}}\equiv\ed \bm{x}/\ed\tau$, with $\tau$ being the affine parameter of the geodesic curves. For the general spacetime \eqref{eq:metr0}, this provides the expression
\begin{equation}
\mL(\bm{x},\dot{\bm{x}})=\frac{1}{2}\left[-f(r)\dot t^2+\frac{\dot r^2}{f(r)}+r^2\left(\dot\theta^2+\sin^2\theta\,\dot\phi^2\right)\right].
    \label{eq:Lagrangian_1}
\end{equation}
Without loss of generality, we confine ourselves to the equatorial plane by setting $\theta=\pi/2$. Hence, the constants of motion can be defined as
\begin{eqnarray}
    && E=f(r)\dot t,\label{eq:E}\\
    && L=r^2\dot\phi,\label{eq:L}
\end{eqnarray}
which correspond, respectively, to the energy and the angular momentum of the test particles. Accordingly, the Hamilton-Jacobi equation $2\mL=0$ results in the following first-order differential equation of motion
\begin{equation}
\dot r^2=E^2-V(r),
    \label{eq:rdot_1}
\end{equation}
in which
\begin{equation}
V(r) = L^2\frac{f(r)}{r^2},
     \label{eq:V}
\end{equation}
is the effective potential for the light rays. Accordingly, using Eq. \eqref{eq:L} we get
\begin{equation}
\left(\frac{\ed r}{\ed\phi}\right)^2=\frac{r^4}{b^2}\left(1-\frac{V(r)}{E^2}\right),
    \label{eq:drdphi}
\end{equation}
where $b=L/E$ is the impact parameter. Photons that reach the maximum effective potential will travel on unstable orbits at a distance $r_p$ from the black hole, defining the radius of the photon sphere. At this point, the orbits are characterized solely by the impact parameter $b$. To determine $r_p$, it is essential first to satisfy the condition $V(r) = E^2$, which corresponds to $\dot{r} = 0 = {\ed r}/{\ed \phi}$ and identifies a turning point. Using this condition, we obtain the impact parameter $b = {r}/{\sqrt{f(r)}}$. To associate this impact parameter with the critical orbits, we also need to satisfy the additional condition $V'(r) = 0$, which determines the maximum of the effective potential and thus identifies $r_p$. Therefore, the critical impact parameter can be expressed as
\begin{equation}
b_c=\frac{r_p}{\sqrt{f(r_p)}},
    \label{eq:bc_0}
\end{equation}
on the photon sphere. The mutual condition $V(r_p)=0=V'(r_p)$ can be solved numerically to show the behavior of $r_p$ as a function of the $\mc$-parameter. In Fig. \ref{fig:rp}, the $\mc$-profile of the photon sphere radius of the black hole has been shown in the positive and negative domains of the $\mc$-parameter. 
{The diagrams indicate that as the $\mc$-parameter increases, the photon sphere shrinks, affecting the final shadow size observed in the sky.} Additionally, larger values of the ratio $M/\lambda$ correspond to larger photon spheres. For $\mc=0$ (the SBH), the well-known values are $r_p=3M$ for the photon sphere radius and $b_c=3\sqrt{3} M$ for the critical impact parameter. 

Now, let us consider an observer at the radial position $r_o$. The shadow radius is then given by the relation \cite{perlick_black_2018}
\begin{equation}
R_s=r_p\sqrt{\frac{f(r_o)}{f(r_p)}}=\sqrt{f(r_o)}\,{b_p}.
    \label{eq:Rs_0}
\end{equation}
For large values of $r_o$, we can then approximate $f(r_o)\approx 1$, and hence, $R_s=b_c$. Such condition is reliable since the observer-source distance for cases like the SMBHs such as M87* and Sgr A* are extremely large and are, respectively, $\approx 16.8\,\rm{Mpc}$ and $\approx 8\, \rm{kpc}$.

\begin{figure}[t]
    \centering
    \includegraphics[width=8cm]{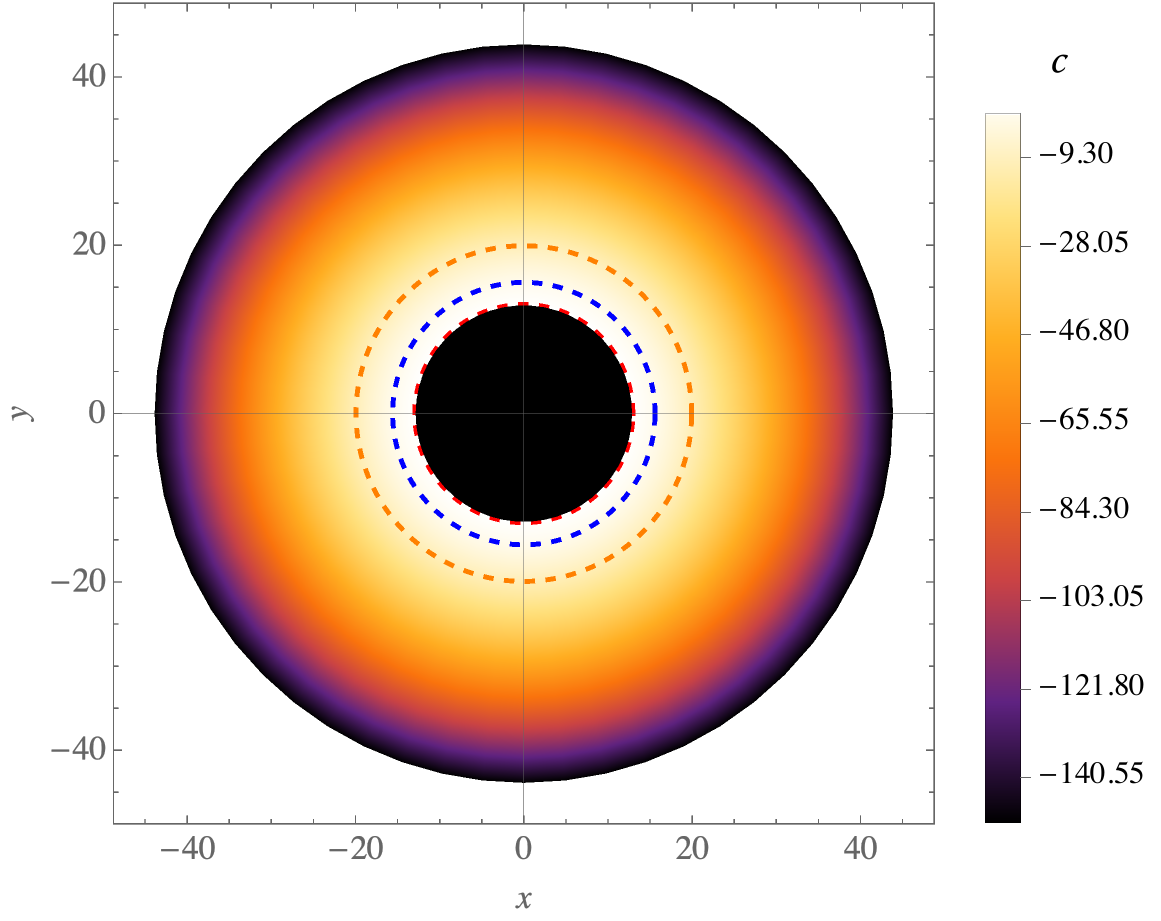}
    \caption{The change in the shadow diameter of a backlit black hole for the case of $M/\lambda=3$, within the domain $-150\leq\mc\leq 3.8$. The orange, blue, and red dashed circles correspond to $\mc=-10$, $\mc=0$ (the SBH), and $\mc=3.7$, respectively. The black disk in the middle, is the lower bound of the shadow.}
    \label{fig:Rs}
\end{figure}

Now, to examine the effect of the scalar hair through the $\mc$-parameter on the shadow diameter of the black hole for an observer located at spatial infinity, we have considered the exemplary case of $M/\lambda=3$ in Fig. \ref{fig:Rs}, within a definite domain of the $\mc$-parameter. 
One can infer from the diagram that, as expected from the $r_p$ profiles, the shadow size in the observer's sky shrinks by a rise in the $\mc$-parameter from negative values to positive values. Hence, the SBH constitutes the lower bound of the shadow size for $\mc<0$ and the upper bound for $\mc>0$. 

At this point, we aim to constrain the model parameters $\mc$ and $\lambda$ using the shadow image data from the EHT observations of M87* and Sgr A* \cite{Akiyama:2019,the_event_horizon_telescope_collaboration_first_2019,the_event_horizon_telescope_collaboration_first_2019-1,do_relativistic_2019,Akiyama:2022,event_horizon_telescope_collaboration_first_2022,gravity_collaboration_mass_2022}. The uncertainties provided in Refs. \cite{kocherlakota_constraints_2021,vagnozzi_horizon-scale_2023} will also be employed. To proceed with the comparisons with the EHT observations, we note that the theoretical shadow diameter of the black hole is given by $d_{\mathrm{sh}}^{\mathrm{theo}} = 2R_s$. 

\begin{figure}[t]
    \centering
    \includegraphics[width=7cm]{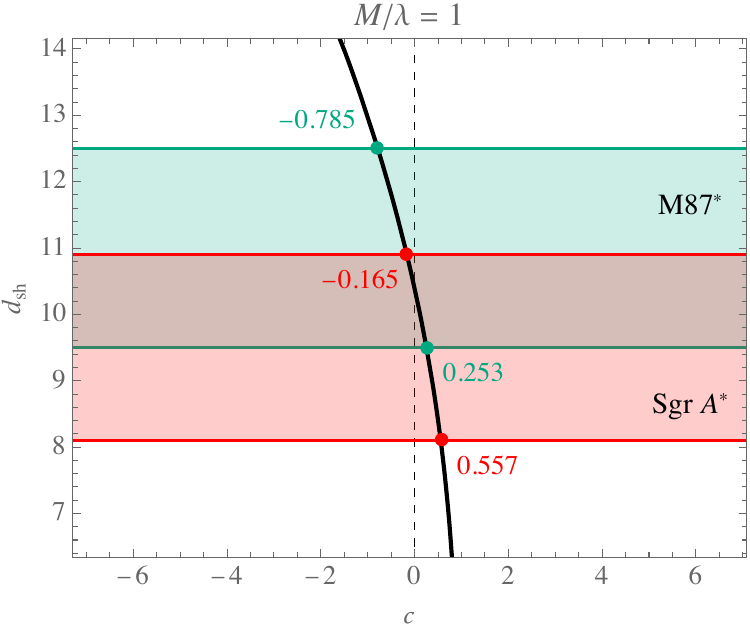}(a)\qquad
    \includegraphics[width=7cm]{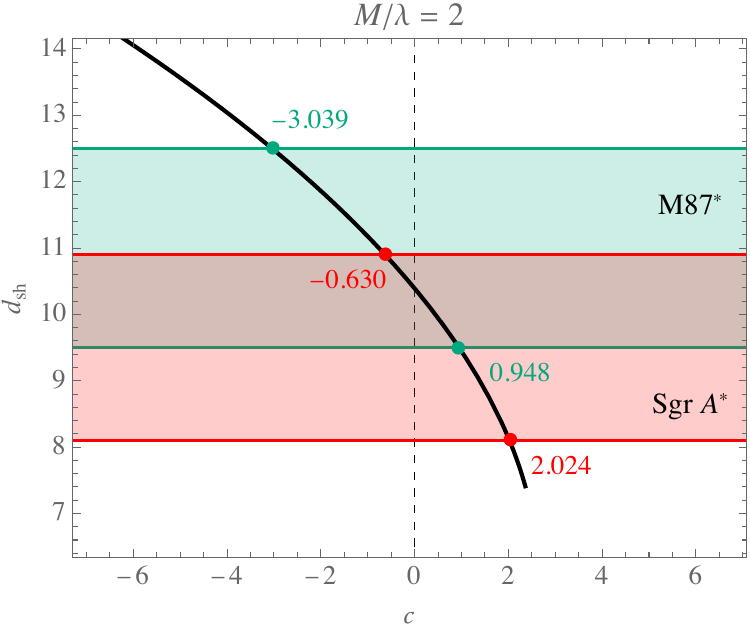}(b)
    \includegraphics[width=7cm]{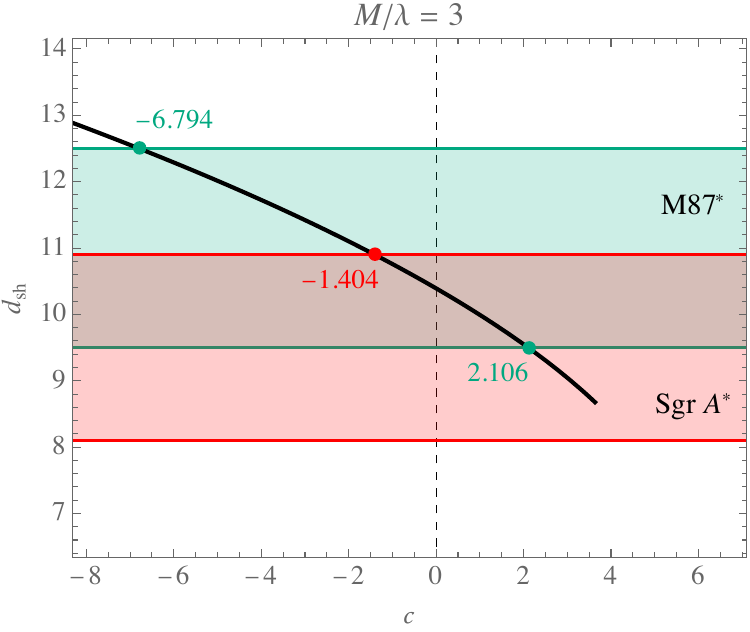}(c)\qquad
    \includegraphics[width=7cm]{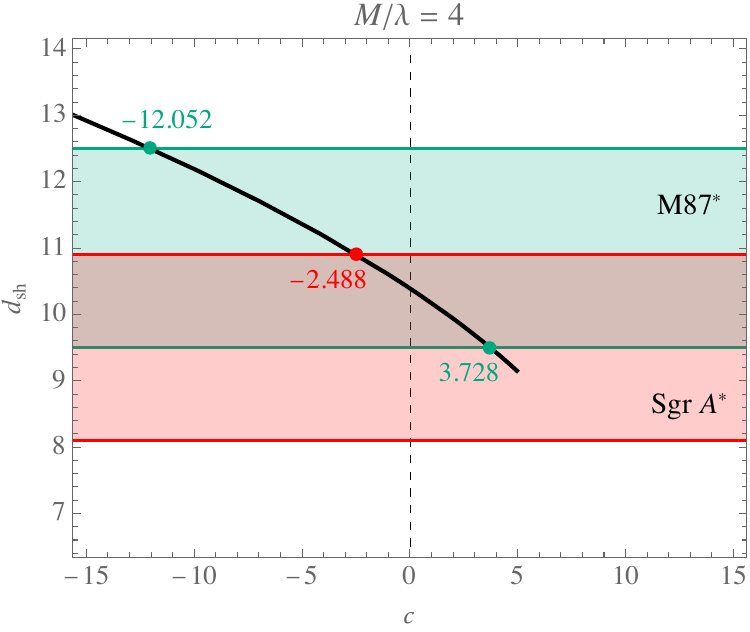}(d)
    \caption{The %rescaled 
    $\mc$-profiles of the 
    %{rescaled: meaning that, respectively, $M$, $2M$, $3M$ and $4M$ are considered as the unit of length; this corresponds to taking into account $M/(\lambda/i)$ with $i=\overline{1,4}$,  as the identifier of the cases $M/\lambda=i$.} 
    theoretical shadow diameter $d_{\mathrm{sh}}^{\mathrm{theo}}$ (black curves) compared with the observed shadow diameters of M87* and Sgr A*, for $M/\lambda={1,2,3,4}$. The points indicate the upper and lower constraints on the $\mc$-parameter for each observation. The gray region is the one in agreement with both observations.}
    \label{fig:EHT_constraints}
\end{figure}

\begin{figure}[H]
    \centering
    \includegraphics[width=7cm]{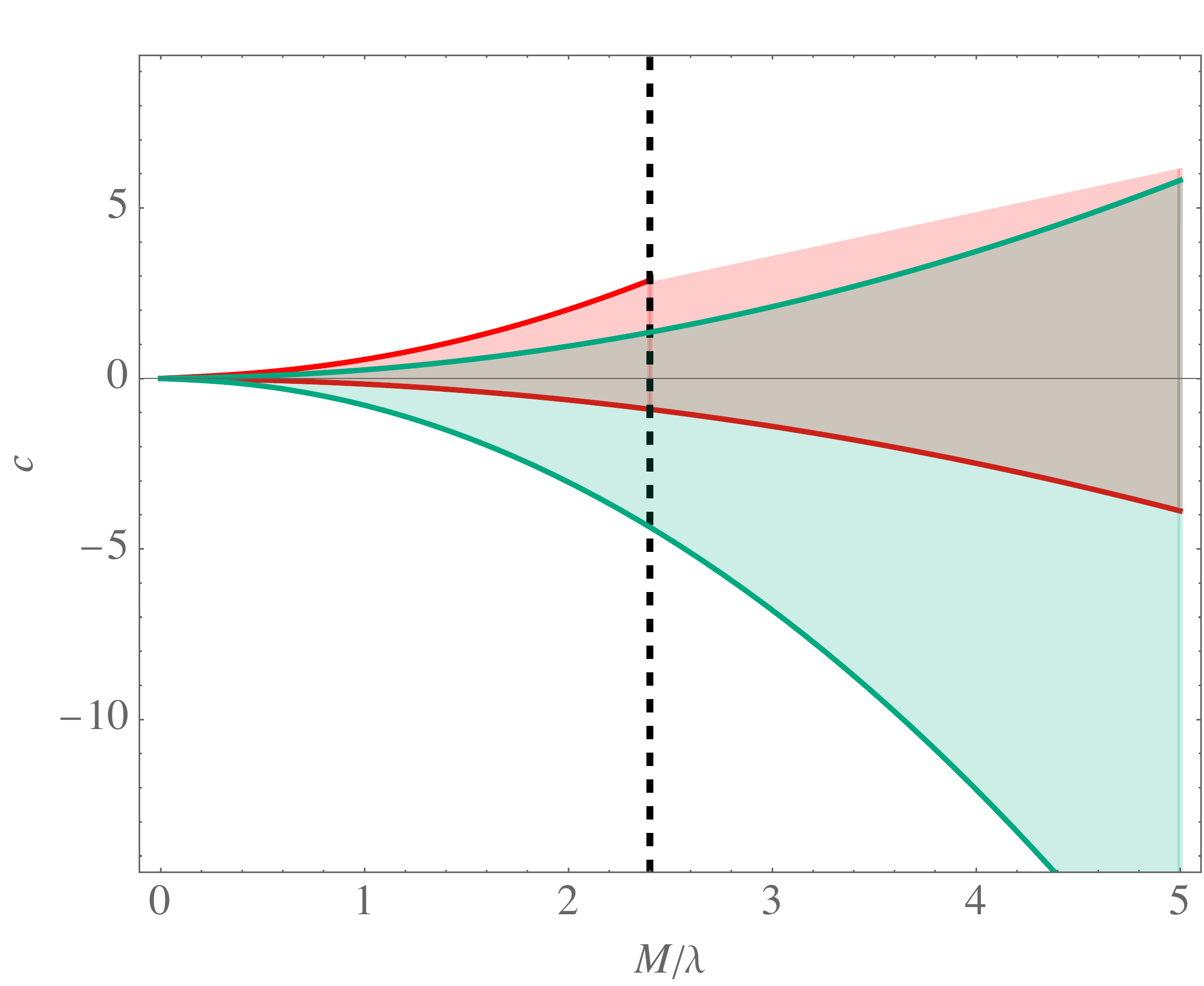} (a)\qquad
    \includegraphics[width=7cm]{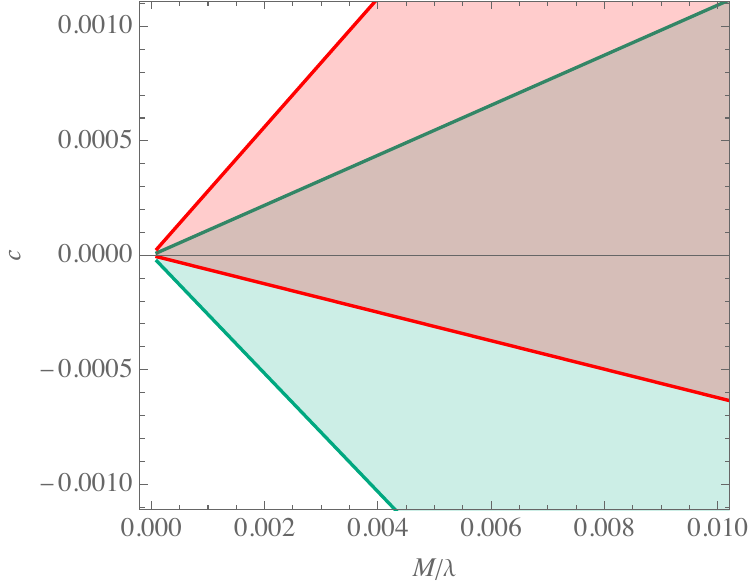} (b)
    \caption{The diagrams show the allowed ranges of $\mc$ and $M/\lambda$ based on EHT observations. The color-coding corresponds to that used in Fig. \ref{fig:EHT_constraints}. The dashed line in panel (a) corresponds to $M/\lambda = 2.4$. After this value, the theoretical lower bound for the diameter size is higher than the lower one constrained by Sgr A* observations. In panel (a), the red region possesses an upper bound for $\mc$ beyond $M/\lambda\approx2.4$ given by $c_*=4M/(\pi \lambda)$. Panel (b) illustrates the parameters' near-zero behavior.}
    \label{fig:crange_gen}
\end{figure}

The diameter of the observed shadows in the recent EHT images of M87* and Sgr A* is calculated using the relation \cite{Bambi:2019tjh}
\begin{equation}
    d_\text{sh} = \frac{D \theta_*}{\gamma M_\odot},    \label{eq:dsh}
\end{equation}
which determines the shadow diameter as observed by an observer positioned at a distance $D$ (in parsecs) from the black hole. Here, $\gamma$ is the mass ratio of the black hole to the Sun. For M87*, $\gamma = (6.5 \pm 0.90) \times 10^9$ at a distance $D = 16.8\,\mathrm{Mpc}$ \cite{Akiyama:2019}, and for Sgr A*, $\gamma = (4.3 \pm 0.013) \times 10^6$ at $D = 8.127\,\mathrm{kpc}$ \cite{Akiyama:2022}. In Eq. \eqref{eq:dsh}, $\theta_*$ is the angular diameter of the shadow, measured as $\theta_*= 42 \pm 3 \,\mathrm{\mu as}$ for M87* and $\theta_* = 48.7 \pm 7 \,\mathrm{\mu as}$ for Sgr A*. Using these values, the shadow diameters can be calculated as $d_{\mathrm{sh}}^{\mathrm{M87*}} = 11 \pm 1.5$ and $d_{\mathrm{sh}}^{\mathrm{SgrA*}} = 9.5 \pm 1.4$. These values are displayed with $1\sigma$ uncertainties for both black holes in Fig. \ref{fig:EHT_constraints}, together with the $\mc$-profile of the theoretical shadow diameter $d_{\mathrm{sh}}^{\mathrm{theo}}$, for four values of the $M/\lambda$ ratio. 
According to the diagrams, as the $M/\lambda$ ratio increases, the profiles tend to extend within the $\mc$ domain, ranging from negative to positive values. Consequently, with an increasing $M/\lambda$ ratio, the observationally respected intervals extend significantly. This results in a considerable rise in the difference between the lower and upper bounds for both M87* and Sgr A*. We can also note that for $M/\lambda=3$ and $M/\lambda=4$, there are no lower bounds for Sgr A* (see Table \ref{tab:1}) as there are no available black hole solutions when the existence lines finish.
\begin{table}[H]
	\centering
	\begin{tabular}{cccccccccccc}
		\toprule
		 &\multicolumn{2}{c}{$M/\lambda=1$}& &\multicolumn{2}{c}{$M/\lambda=2$}& &\multicolumn{2}{c}{$M/\lambda=3$}& &\multicolumn{2}{c}{$M/\lambda=4$}\\
		\cmidrule{2-3} \cmidrule{5-6} \cmidrule{8-9} \cmidrule{11-12}
		
		{} & {upper} & {lower} &  {} & {upper} & {lower}  & {} & {upper} & {lower}&  {} & {upper} & {lower}\\
		\midrule
		$\text{M87*}$ & $-0.78546$ & $0.25278$ & & $-3.03896$ & $0.94822$ & & $-6.79433$ & $2.10641$ & & $-12.05180$ & $3.72775$\\
		$\text{Sgr A*}$ & $-0.16481$ & $0.55745$ & & $-0.62962$ & $2.02362$ & & $-1.40409$ & - & & $-2.48832$ & -\\
		\bottomrule
	\end{tabular}
 \caption{The allowable $\mc$-parameter values, obtained from the curves in Fig. \ref{fig:EHT_constraints}, corresponding to the black hole shadow diameter $d_{\rm{sh}}$ that aligns with the EHT observations of M87* and Sgr A* within the $1\sigma$ confidence intervals.}
 \label{tab:1}
\end{table}
The absence of a lower theoretical bound for Sgr A* shadow diameter begins at a specific value of $M/\lambda$, which correlates with the changes in the photon sphere radius, as shown in Fig. \ref{fig:rp}(b). Beyond this value, the lowest possible theoretical shadow diameter is higher than the lower bound constrained by $1\sigma$ confidence intervals for Sgr A*. This critical value of $M/\lambda$ is highlighted in Fig. \ref{fig:crange_gen}(a), where the range of acceptable values for $\mc$ and $M/\lambda$ that align with the EHT observations is depicted. After this critical value, the upper bound for $\mc$ compatible with Sgr A* observations is governed by $\mc_*=4M/(\pi \lambda)$.
%

%%%%%%%%%%%%%%sec:5

\section{Spherically infalling accretion and the disk's images}\label{sec:infall}

In this section, we continue our study by examining the shadow cast by the black hole as it spherically accretes radiative gas, forming a thin emission disk \cite{bambi_can_2013}. The observed intensity of the emission is given by
\begin{equation}
I_{\rm{obs}} = \int_{\bm{\gamma}} \mathcal{R}^3 {J}(\nu_{\rm{e}}) \,\ed I_{\rm{prop}},
    \label{eq:I_obs_infalling_0}
\end{equation}
where the integration is performed over the null geodesic congruence $\bm{\gamma}$. In this expression, $\mathcal{R}$ denotes the redshift factor, $\nu_{\rm{e}}$ is the frequency of the emitted photons from the accretion disk, and $\ed I_{\rm{prop}}$ represents the infinitesimal proper length. The emission per unit volume in the emitter's rest frame is described by
\begin{equation}
{J}(\nu_{\rm{e}}) \propto \frac{\delta(\nu_{\rm{e}} - \nu_{\rm{f}})}{r^2},
    \label{eq:j(nu)_0}
\end{equation}
where $\nu_{\rm{f}}$ is the monochromatic emission frequency in the rest frame and $\delta$ is the Dirac delta function. The redshift factor is given by
\begin{equation}
\mathcal{R} = \frac{\Pi_\mu u_{\rm{o}}^{\mu}}{\Pi_{\nu} u_{\rm{e}}^\nu},
    \label{eq:redshift_factor_0}
\end{equation}
with $\bm{u}_{\rm{o}}$ and $\bm{u}_{\rm{e}}$ representing the four-velocities of a distant static observer and the infalling matter, respectively. Specifically, $u_{\rm{o}}^{\mu} = (1,0,0,0)$, and in the spacetime of the black hole, we have
\begin{equation}
u_{\rm{e}}^{\mu} = \left(\frac{1}{f(r)}, -\sqrt{1 - f(r)}, 0, 0\right).
    \label{eq:u_e}
\end{equation}
The covector $\bm{\Pi}$ in Eq.~\eqref{eq:redshift_factor_0} represents the four-momentum of the emitted photons. Since the accretion is purely radial, we need only compute the temporal and radial components of $\bm{\Pi}$, leading to
\begin{equation}
\frac{\Pi_r}{\Pi_t} = \pm \frac{1}{f(r)} \sqrt{1 - f(r) \frac{b^2}{r^2}},
    \label{eq:conj_moment_new}
\end{equation}
where the $\pm$ sign indicates whether photons are approaching or receding from the black hole. Consequently, the redshift factor can be rewritten as
\begin{eqnarray}
\mathcal{R} &=& \left(u_{\rm{e}}^t + \frac{\Pi_r}{\Pi_t} u_{\rm{e}}^r \right)^{-1} \nonumber \\
&=& \left[\frac{1}{f(r)} \pm \sqrt{\left(\frac{1}{f(r)} - 1\right) \left(\frac{1}{f(r)} - \frac{b^2}{r^2}\right)}\right]^{-1},
    \label{eq:redshift_factor_1}
\end{eqnarray}
and the infinitesimal proper length is given by
\begin{equation}
\ed I_{\rm{prop}} = \Pi_\mu u_{\rm{e}}^\mu \ed\tau = \frac{\Pi_t}{\mathcal{R} |\Pi_r|} \ed r.
    \label{eq:Iprop}
\end{equation}
Thus, Eq.~\eqref{eq:I_obs_infalling_0} becomes
\begin{equation}
I_{\rm{obs}} \propto \int_{\bm{\gamma}} \frac{\mathcal{R}^3}{r^2} \frac{\Pi_t}{|\Pi_r|} \ed r,
    \label{eq:I_obs_infalling_1}
\end{equation}
and the observed intensity is obtained by integrating this expression over all frequencies. This approach enables us to analyze the brightness of the infalling accretion and the shadow of the black hole. To visually demonstrate the impact of scalar hair on the observational image of the black hole, we begin by fixing some values for the black hole parameters. Based on the analysis from the previous section, we select representative $\mc$ values whose shadow diameters agree with M87* and Sgr A* observed shadows. Specifically, we will focus on cases that lie between the $\mc$ upper bound for M87* and the $\mc$ lower bound for Sgr A* (see Table \ref{tab:1}) for some of the $M/\lambda$ ratios used previously in Fig. \ref{fig:EHT_constraints}.
\begin{figure}[h]
    \centering
 \includegraphics[width=5.4cm]{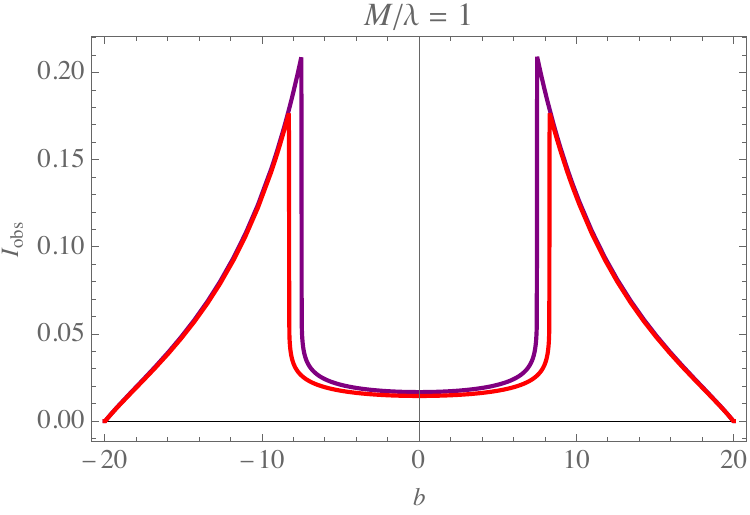} (a)
     \includegraphics[width=5.4cm]{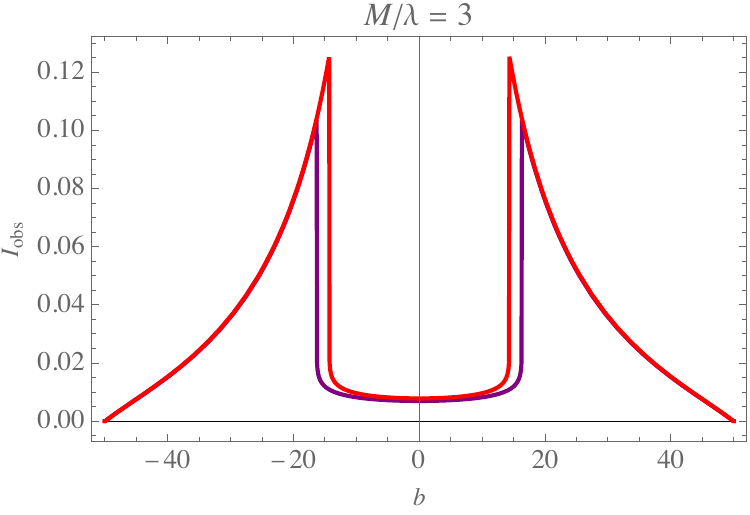} (b)
     \includegraphics[width=5.4cm]{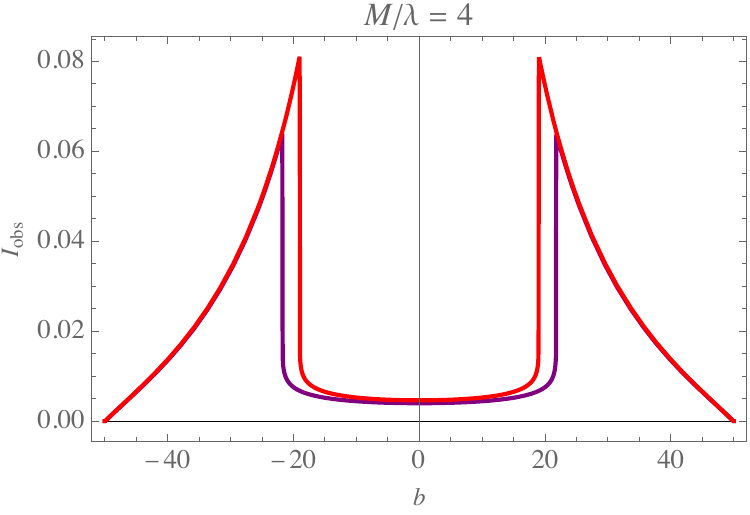} (c)
    \caption{The $b$-profiles of the observed intensity for spherical accretion. In all panels, the purple curve corresponds to negative values for the $\mc$-parameter, while the red curves consider the positive values. Hence the diagrams correspond to (a) $\mc=(-0.1,0.2)$, (b) $\mc=(-1.3,2)$, and (c) $\mc=(-2.4,3.7)$.}
    \label{fig:I_obs_infall}
\end{figure}
As illustrated in Fig. \ref{fig:I_obs_infall}, for all chosen values of the $\mc$-parameter, as the impact parameter increases, the specific observed intensity increases until it reaches a peak\footnote{A \textit{Mathematica} notebook specifically prepared for this purpose was employed. This notebook is a modified version of the one used in Ref. \cite{okyay_nonlinear_2022}, and has also been used in Ref. \cite{fathi_observational_2023}}. Beyond this peak, the intensity drops sharply and approaches zero. The diagrams also indicate that, for the case of $M/\lambda=1$, the peaks are higher when $\mc<0$ compared to when $\mc>0$, which is in stark contrast to the other cases of $M/\lambda$. Additionally, as the $M/\lambda$ values increase, the minimum and maximum intensities decrease. This trend is also reflected in the accretion disk images presented in Fig. \ref{fig:shadow_infall}. 
\begin{figure}[t]
    \centering
    \includegraphics[width=7cm]{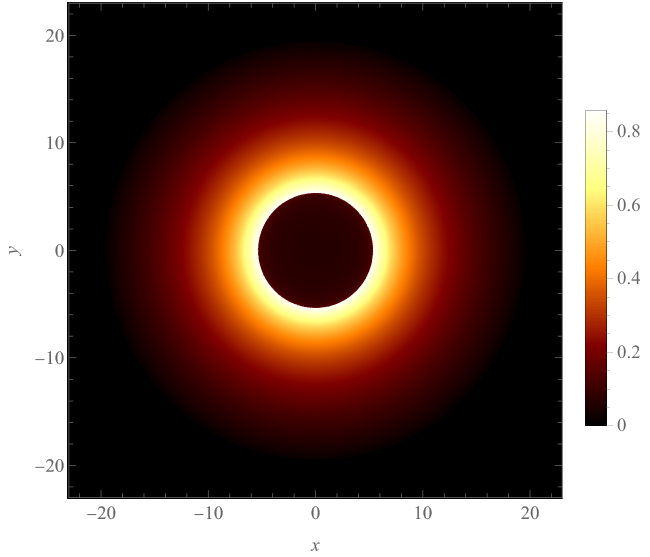}\qquad
     \includegraphics[width=7cm]{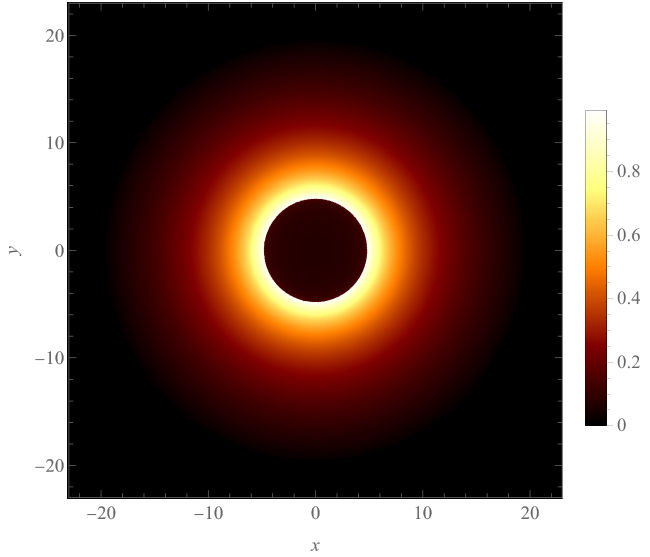}~(a)
     \includegraphics[width=7cm]{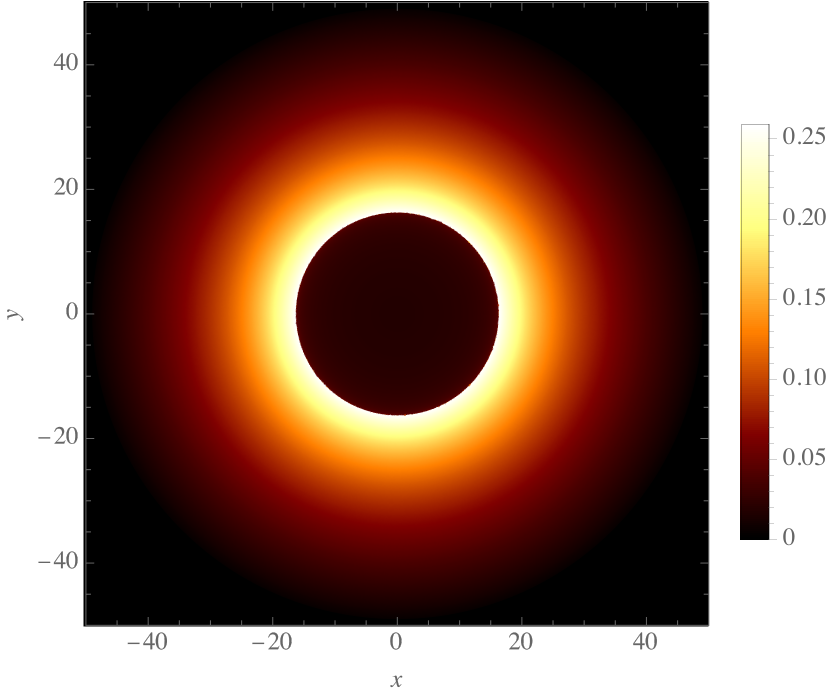}\qquad
     \includegraphics[width=7cm]{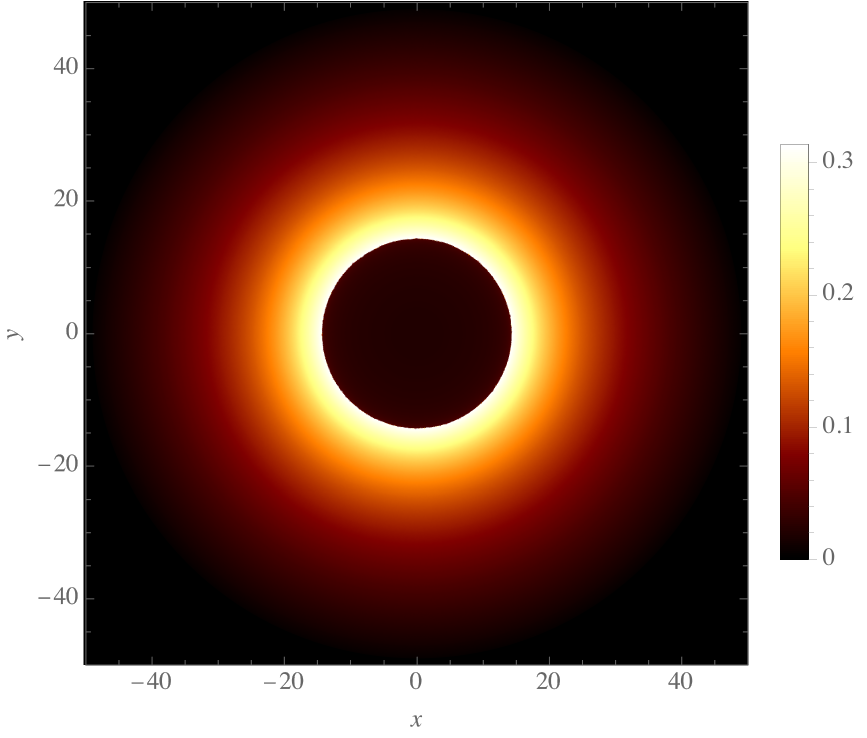}~(b)
      \includegraphics[width=7cm]{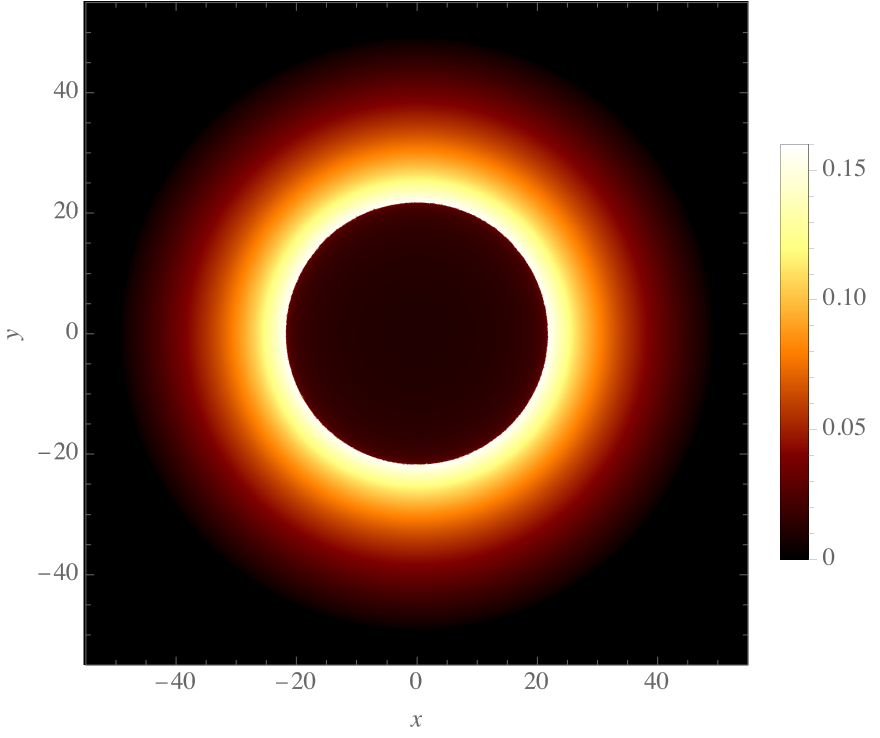}\qquad
      \includegraphics[width=7cm]{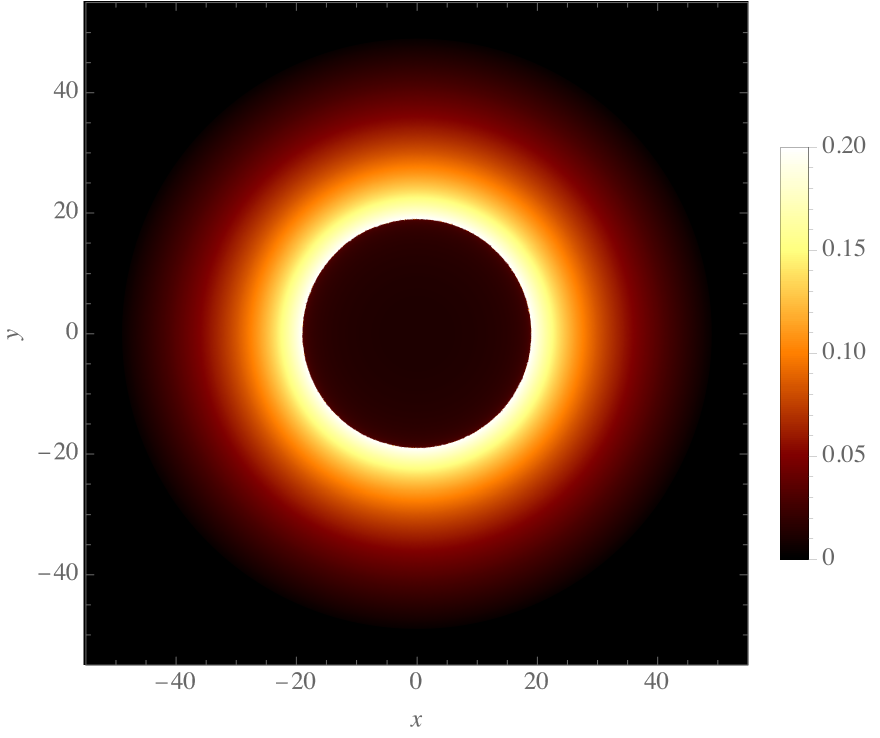}~(c)
    \caption{The images illustrate the disk and black hole silhouette with infalling spherical accretion. The rows correspond to (a) $M/\lambda=1$, (b) $M/\lambda=3$, and (c) $M/\lambda=4$. The left panels depict cases with $\mc < 0$, while the right panels represent cases with $\mc > 0$, as detailed in Fig. \ref{fig:I_obs_infall}.}
    \label{fig:shadow_infall}
\end{figure}
As seen in the diagrams, an increase in $M/\lambda$ leads to an enlargement of the silhouette. For each $M/\lambda$ value, the silhouette is larger when $\mc>0$ compared to $\mc<0$. However, for $M/\lambda=1$, the accretion disk is brighter when $\mc<0$ than when $\mc>0$, contrasting with the other two cases of $M/\lambda$. These observations underscore the sensitivity of black hole images with infalling accretion to the scalar hair, as well as the other two black hole parameters.

%%%%%%%%%%%%%%%
\section{Summary and conclusions}\label{sec:conclusions}

In this article, we explored a black hole spacetime endowed with primary scalar hair, examining its thermodynamic properties, null geodesics, and observational signatures. Our results revealed a profound influence of scalar hair on the black hole's physical properties, particularly its shadow and observable characteristics. We began by examining the thermodynamic behavior of the black hole, where we observed that the presence of primary scalar hair introduces notable modifications to the black hole's temperature, entropy, and specific heat. In particular, the primary hair generates intermediate-size black hole branches that are locally stable and not close to Schwarzschild black in the parameters space. Namely, we have found that locally stable black holes exist only for
\begin{equation*}
\mc\lessapprox-9.097\quad \text{and} \quad \mc\gtrapprox1.319\ .
\end{equation*}
In our analysis of null geodesics, we discovered that scalar hair significantly influences the shadow cast by the black hole. {We find that the shadow size decreases as the parameter $\mc = \eta q^4$ is increased from negative to positive values. Thus, the effect of primary hair on the shadow size, within a given theory with fixed $\eta$, can be explained as follows: in the negative domain of $\mc$, where $\eta < 0$, a larger scalar hair leads to a larger shadow. Conversely, in the positive domain of $\mc$, where $\eta > 0$, a larger scalar hair results in a smaller shadow (see Fig. \ref{fig:Rs}).} This observation is particularly relevant given the recent imaging data from the EHT. By applying the EHT constraints to our model, we found that the scalar hair parameter can be bounded within certain limits, depending on the black hole’s mass-to-coupling ratio. Our results demonstrate that the shadow size is highly sensitive to both the magnitude and sign of the scalar hair and other black hole parameters. This sensitivity underscores the importance of accurately determining these parameters to effectively constrain the presence of scalar hair through observational data. Remarkably, the family of theories given by $\lambda$ and $\eta$ always contain black holes with primary hair in agreement with both EHT observed shadows, and such range on the primary hair increases as the mass of the black hole is higher.

We further explored the effects of scalar hair on photon orbits near the black hole, finding that it alters the effective potential, thereby modifying the structure of the orbits that define the black hole’s shadow. These modifications have direct implications for the black hole’s observational appearance. Our simulations of thin accretion disk images revealed that scalar hair affects both the brightness distribution and the shape of the disk. Notably, for certain values of the scalar hair parameter, the accretion disk images exhibit an asymmetry in brightness, with a noticeable difference between positive and negative scalar hair values. For example, in the case where $M/\lambda=1$, the disk is brighter for $\mc<0$ than for $\mc>0$, which contrasts with other cases and highlights the complex interplay between the scalar hair and black hole parameters. These results suggest that high-resolution observations, such as those expected from future EHT campaigns, could potentially detect these subtle differences, providing a new avenue for probing scalar hair. The comparative analysis across different $M/\lambda$ ratios revealed distinct behaviors in the peak intensities and shadow sizes for positive versus negative scalar hair. In particular, the anomaly observed at $M/\lambda=1$, where the intensity peaks for $\mc<0$ are higher than those for $\mc>0$, suggests a unique behavior of the black hole spacetime in this specific regime. Such complexities in the behavior of scalar hair could be crucial for distinguishing between various black hole models, especially in the context of interpreting observational data.

Our study provides compelling evidence that primary scalar hair profoundly impacts the thermodynamic and observational properties of black holes. Constraints from local thermodynamic stability and constraints obtained by observations can be intersected. In doing so, we found that, unlike general relativity, the primary hair allows locally thermodynamically stable black holes whose shadow sizes are compatible with current EHT observations. The sensitivity of black hole shadow and accretion disk images to scalar hair, as constrained by EHT data, opens up new possibilities for testing the existence of such hair through direct astronomical observations. As the resolution and sensitivity of black hole imaging techniques continue to advance, the influence of scalar hair could become a detectable feature, offering a unique probe into the nature of black holes and the fundamental theories that describe them.

Certainly, more research is needed to extend these analyses to rotating black holes and explore how the theory's parameters affect circularity. Additionally, further observational efforts should aim to refine the constraints on scalar hair using more extensive datasets from EHT and other upcoming interferometric arrays. {It is important to note that, in this work, we have used the black hole shadow size as the main tool to constrain the black hole parameters. As a result, no specific constraint has been placed on the parameter $\lambda$. To refine the observational constraints further, it is also essential to consider additional experimental tests, such as the period and periapsis shift in the orbits of S-stars near the center of the Milky Way. Incorporating such new experimental data could provide a more comprehensive framework for testing the black hole model and constraining its parameters.} These endeavors hold the potential to significantly enhance our ability to test alternative theories of gravity and deepen our understanding of black hole physics.

%%%%%%%%%%%Acknowledgements 
\section*{Acknowledgements}
 C.E. is funded by Agencia Nacional de Investigación y Desarrollo (ANID) through Proyecto Fondecyt Iniciación folio 11221063, Etapa 2024. M.F. is supported by Universidad Central de Chile through project No. PDUCEN20240008.

%%%%%%%%%%%%%%%%appendices
%\appendix
%%%%%%%%%%%%%References
\bibliographystyle{ieeetr}
\bibliography{biblio.bib}

\end{document}